\documentclass[pra,aps,10pt,superscriptaddress,twocolumn,floatfix]{revtex4-1}  
\usepackage{graphicx,color}
\usepackage{amsmath,amssymb,bm}
\usepackage{braket}		

\DeclareMathOperator{\Tr}{Tr}

\usepackage[plainpages=false,pdfpagelabels,colorlinks=true,linkcolor=red,urlcolor=blue,citecolor=blue,pdftitle={Titl}e,pdfauthor={},pdfdisplaydoctitle=true,pdfduplex=DuplexFlipLongEdge]{hyperref}

\definecolor{darkred}{rgb}{0.90,0.2,0.2}
\definecolor{darkgreen}{rgb}{0,0.60,.2}
\definecolor{darkblue}{rgb}{0.1,0.3,1}
\definecolor{grey}{cmyk}{0,0,0,0.25}
\definecolor{orange}{cmyk}{0,0.6,0.8,0}

\begin{document}
\title{Information measures for local quantum phase transitions: \\
Lattice bosons in a one-dimensional harmonic trap}

\author{Yicheng Zhang}
\affiliation{Department of Physics, The Pennsylvania State University, University Park, Pennsylvania 16802, USA}
\author{Lev Vidmar}
\affiliation{Department of Theoretical Physics, J. Stefan Institute, SI-1000 Ljubljana, Slovenia}
\affiliation{Department of Physics, University of Ljubljana, SI-1000 Ljubljana, Slovenia}
\author{Marcos Rigol}
\affiliation{Department of Physics, The Pennsylvania State University, University Park, Pennsylvania 16802, USA}

\begin{abstract}
We study ground-state quantum entanglement in the one-dimensional Bose-Hubbard model in the presence of a harmonic trap. We focus on two transitions that occur upon increasing the characteristic particle density: the formation of a Mott-insulating domain with site occupation one at the center of the trap (lower transition) and the emergence of a superfluid domain at the center of the Mott-insulating one (upper transition). These transitions generate discontinuities in derivatives of the total energy and have been characterized by local (nonextensive) order parameters, so we refer to them as local quantum phase transitions. We show that a second derivative of the total energy is continuous with a kink at the lower transition, and that it is discontinuous at the upper transition. We also show that bipartite entanglement entropies are order parameters for those local quantum phase transitions. We use the density-matrix renormalization group and show that the transition points extracted from entanglement measures agree with the predictions of the local density approximation in the thermodynamic limit. We discuss how to determine the transition points from results in small systems, such as the ones realized in recent optical lattice experiments that measured the second-order Renyi entanglement entropy.
\end{abstract}
\maketitle


\section{Introduction} \label{sec1}

As a fundamental property of quantum mechanics, entanglement and its measures contain information of nonlocal correlations of quantum states~\cite{horodecki_09}. For many-body lattice systems, entanglement measures provide powerful tools to characterize their key properties~\cite{amico_fazio_08, eisert_cramer_10, laflorencie_16}. Entanglement has also provided new perspectives on quantum phase transitions~\cite{osterloh_amico_2002, osborne_nielsen_02, vidal_03, calabrese_04}, quantum dynamics and thermalization~\cite{calabrese_2005, chiara_2006, Daley_2012, Schachenmayer_2013, kim_2013}, and topological order~\cite{hamma_05, kitaev_06, levin_06}. In addition to extensive theoretical interest recent experimental progress with quantum gases, involving quantum interference~\cite{islam_ma_15, kaufman_tai_16} and randomized measurements~\cite{brydges_19}, have made possible studies of quantum entanglement via the second-order Renyi entanglement entropy.

In this paper, we study quantum entanglement in the ground state of bosons in one-dimensional lattices in the presence of a harmonic trapping potential, which models the experimental system in Refs.~\cite{islam_ma_15, kaufman_tai_16}. Trapped lattice systems have been studied systematically~\cite{batrouni_rousseau_02, kashurnikov_prokofev_02, rigol_muramatsu_03, rigol_muramatsu_04may, kollath_schollwoeck_04, rigol_muramatsu_04b, hooley_quintanilla_04, bergkvist_henelius_04, wessel_alet_04, rigol_muramatsu_04sept, rigol_muramatsu_05july, rey_pupillo_05, batrouni_krishnamurthy_08, rigol_batrouni_09, campostrini_vicari_09, pollet_prokofev_10, Spielman_10, campostrini_vicari_10b, campostrini_vicari_10c, mahmud_duchon_11, pollet_12, ceccarelli_torrero_13, angelone_campostrini_14, zhang_18} since their early experimental realizations with ultracold atoms~\cite{jaksch_98, greiner02, greiner_mandel_02b, bloch08, cazalilla_citro_review_11}. A special feature of trapped lattice bosons, different from their homogeneous counterparts, is the absence of a tradition quantum phase transition paradigm because the Mott-insulating and superfluid regions can coexist \cite{batrouni_rousseau_02, kashurnikov_prokofev_02, rigol_batrouni_09}. As a result, the formation of a Mott-insulating domain at the center of the trap does not result in a vanishing compressibility, which is a global order parameter for the homogeneous system. 

Instead, the formation of Mott-insulating domains has been identified using local quantities, such as the local compressibility and the fluctuations of the site occupations~\cite{batrouni_rousseau_02, wessel_alet_04, rigol_batrouni_09}, and we refer to them as local quantum phase transitions. For trapped spinless fermions~\cite{rigol_muramatsu_04b, hooley_quintanilla_04, zhang_18}, the emergence of a band-insulating domain at the center of the trap produces a kink in the second derivative of the total energy as a function of the characteristic density $\rho = N/R$, where $N$ is the total number of particles and $R$ characterizes the curvature of the trap~\cite{zhang_18} [see Eq.~\eqref{H_BH}]. This motivates our use of the term quantum phase transition in trapped systems.

Here, we study two local quantum phase transitions that occur in the Bose-Hubbard model~\cite{Fisher_1989} in the presence of a harmonic trap: the formation of the $n=1$ Mott-insulating domain at the center of the trap (referred to as the {\it lower transition}) and the emergence of a superfulid domain with $n>1$ at the center of the $n=1$ Mott-insulating one (referred to as the {\it upper transition}), where $n$ denotes the site occupations. Both transitions are driven by increasing the characteristic density $\rho$. We show that the second derivative of the total energy with respect to the characteristic density is continuous with a kink at the lower transition (as found in Ref.~\cite{zhang_18} for the formation of a band-insulating domain for spinless fermions), and that it is discontinuous at the upper transition. The goal of our paper is to characterize those transitions by means of ground-state entanglement entropies, which we compute using density-matrix renormalization group (DMRG) calculations~\cite{white_92, schollwoeck_05, schollwoeck_11}.

Quantum entanglement in the Bose-Hubbard model has been widely explored in homogeneous~\cite{buonsante_07,deng_11, ejima_12,pino_12, alba_12, alba_13,frerot_16}, as well as disordered~\cite{deng_2013,Goldsborough_2015} systems. In the homogeneous case, an area law scaling (with a logarithmic correction) of the entanglement entropy with the subsystem size was demonstrated for both the superfluid and the Mott-insulating phases with a singularity at the transition point~\cite{frerot_16}. For a gapped Mott-insulating phase in one dimension, the relationship between the entanglement spectrum and the interaction strength was analyzed using perturbation theory~\cite{alba_12}. On the other hand, for trapped systems, previous studies of quantum entanglement mainly focused on free models~\cite{Campostrini_2010, Calabrese_2011, vicari12, Calabrese_2015, dubail_stephan_17, dubail_stephan_17b, eisler_bauernfeind_17, zhang_18, tonni_rodriguez_18, murciano_ruggiero_19}.

In a recent study~\cite{zhang_18}, we showed that the ground-state entanglement entropy of trapped spinless fermions in one-dimensional lattices serves as an order parameter for the local quantum phase transition that occurs when a band-insulating domain forms at the center of the trap. Here, we demonstrate that the same is true for what we refer to as the lower and upper transitions in the trapped Bose-Hubbard model. We show that about the lower transition there exists a scaling function that is a universal function of the characteristic density $\rho$ for any given value of the strength of the on-site repulsion, similar to what was previously found for spinless fermions~\cite{zhang_18}. While the lower transition is characterized by a smooth decrease in the entanglement entropy, the upper transition is characterized by a sharp increase. We outline a scaling analysis that enables one to determine the critical characteristic density for the latter transition in the thermodynamic limit. 

The use of the entanglement entropy for studying both transitions allows us to determine the critical values of the characteristic density much more accurately than using local properties~\cite{rigol_batrouni_09}. We also study the entanglement entropy for small numbers of particles, of the order accessible in current experimental setups~\cite{islam_ma_15}. We carry out extrapolations for those particle numbers and show that the critical characteristic densities can be extracted in a robust way.

The presentation is organized as follows: In Sec.~\ref{sec2}, we introduce the model and use the local density approximation (LDA) to determine the critical characteristic densities in the thermodynamic limit. We study the ground-state bipartite entanglement entropy upon the formation of the $n=1$ Mott insulator (lower transition) and the emergence of the $n>1$ superfluid (upper transition) in Secs.~\ref{sec3_1} and~\ref{sec3_2}, respectively. A summary of our results in presented in Sec.~\ref{sec_conclusion}.

\section{Model and local density approximation} \label{sec2}

We study the one-dimensional Bose-Hubbard model in the presence of an external harmonic confining potential. The Hamiltonian can be written as
\begin{align}\label{H_BH}
\hat H =&-t \sum_{i=1}^{L-1}(\hat b^\dagger_i \hat b^{}_{i+1}+\text{H.c.})+\frac{U}{2}\sum_{i=1}^{L}\hat n_i(\hat n_i-1)\nonumber\\ &+\frac{ta^2}{R^2}\sum_{i=1}^{L}\left(i-\frac{L+1}{2}\right)^2\hat n_i\,,
\end{align}
where $\hat b^\dagger_i$ ($\hat b_i$) is the creation (annihilation) operator of a boson at site $i$, $\hat n_i=\hat b^\dagger_i \hat b^{}_i$, $t$ is the hopping amplitude, and $U$ is the strength of the on-site repulsion. The total number of lattice sites is $L$ (taken to be even), $R$ determines the curvature of the harmonic trap, and we set the trap center to be at $x_0=(L+1)a/2$. In a system with $N$ particles, the characteristic density $\rho$ is defined as $\rho=N/R$~\cite{rigol_muramatsu_03, rigol_muramatsu_04b}. We use $\rho$ and the strength of the on-site repulsion $U$ to characterize the properties of the trapped system. In what follows, we set the hopping amplitude $t=1$ as the unit of energy and the lattice spacing $a=1$ as the unit of distance.

The results reported throughout this paper are from DMRG simulations carried out using the \textsc{itensor} library~\cite{itensor}. We set the maximum bond dimension for the matrix product states to be $3200$, the truncation error cutoff to be $10^{-12}$, and the maximum number of bosons per site to be $6$. The convergence criteria is set such that the energy difference between two consecutive sweeps is smaller than $10^{-11}$.

In homogeneous systems ($R^{-2}=0$) [see Eq.~(\ref{H_BH})], the Bose-Hubbard model is known to exhibit a Mott-insulating phase for $U>U^{n}_c$ when the particle occupation $n$ is an integer number, and a superfluid phase otherwise~\cite{Fisher_1989}. In trapped systems, superfluid and Mott-insulator domains can coexist. One can understand this within the local density approximation (LDA)~\cite{bergkvist_henelius_04, batrouni_krishnamurthy_08}. Within the LDA, one replaces the confining potential term in Eq.~(\ref{H_BH}) by an effective local chemical potential,
\begin{equation}\label{LDA}
\mu(x)=\mu_0-\frac{x^2}{R^2}\,,
\end{equation}
where $x$ is the distance from the center of the trap and $\mu_0$ is the chemical potential at the center of the trap. The region about site $x$ is thought of as a homogeneous system with chemical potential $\mu(x)$. Depending on $\mu(x)$ and $U$, one can then have different local ``phases'' (domains) in different regions of the trap.

\begin{figure}[!t]
\begin{center}
\includegraphics[width=0.99\columnwidth]{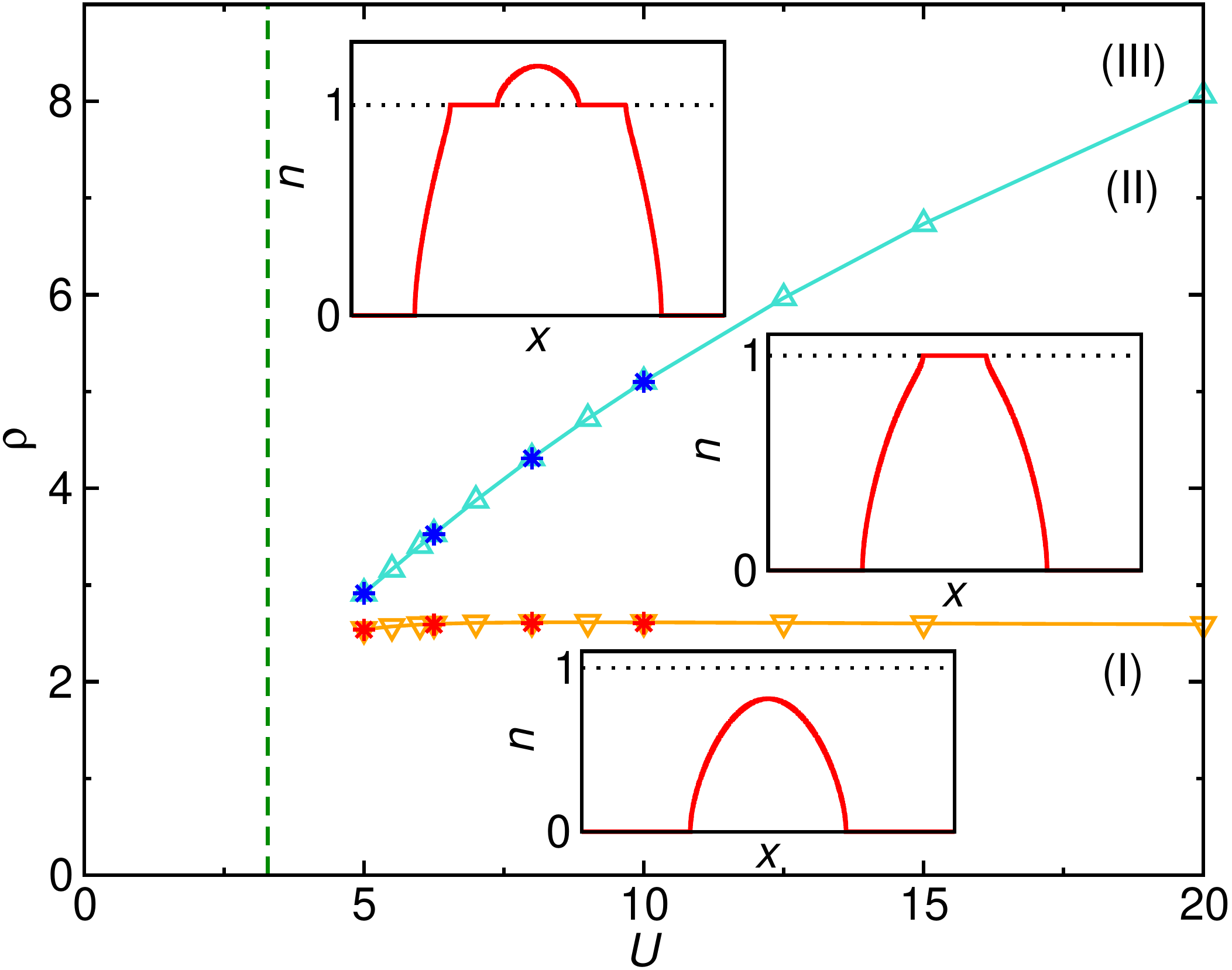}
\caption{State diagram of the trapped Bose-Hubbard model in Eq.~(\ref{H_BH}) (see also Ref.~\cite{rigol_batrouni_09}). Labels for $U>U_c^{n=1}$ indicate the states of the trapped system exemplified in the corresponding insets: (I) $n<1$ superfluid profile, (II) $n=1$ Mott-insulator at the center of the trap surrounded by $n<1$ superfluid regions, and (III) $n>1$ superfluid region at the center of the $n=1$ Mott insulating domain that, in turn, is surrounded by $n<1$ superfluid regions. The critical values of the characteristic density $\rho$ for the transition from (I) to (II) ($\rho^l_c$, lower transition, down-triangles) and from (II) to (III) ($\rho^u_c$, upper transition, up-triangles) were obtained using the LDA [Eq.~(\ref{rho_c})] from calculations in homogeneous systems with size $L_0=200$. We also show results (stars) for the extrapolated ($L_0\to\infty$) LDA predictions, see Fig.~\ref{lda}. The vertical dashed line indicates the critical value $U^{n=1}_c=3.28$ for the formation of $n=1$ Mott insulator in the homogeneous system~\cite{carrasquilla_manmana_13, Ejima_2011}.} 
\label{statediagram}
\end{center}
\end{figure}

Hence, within the LDA, in order to compute the site-occupation profiles in a harmonic trap $n(x)=n[\mu(x),U]$, all one needs to do is to compute the site occupations $n(\mu,U)$ in a homogeneous system. Using DMRG, we compute $\mu(n,U)$ in finite homogeneous systems with $L_0$ sites and $N$ particles ($n=N/L_0$) and open boundary conditions. The chemical potential is obtained as $\mu(n,U)=E(N,U)-E(N-1,U)$, where $E(N,U)$ is the ground-state energy of a system with $N$ particles and on-site repulsion $U$~\cite{Ejima_2011}. We then perform an interpolation (using a Hermite interpolation method with a cubic order) to make $\mu(n,U)$ a continuous function and to establish the inverse relation $n(\mu,U)$. The next step is to determine the offset $\mu_0$ in Eq.~(\ref{LDA}), i.e., the chemical potential at the trap center. The lower (upper) transition occurs when $\mu_0$ just reaches (leaves) the $n=1$ Mott-insulating phase, $\mu_0=\mu_{l}(U)=E(L_0, U)-E(L_0-1, U)$ [$\mu_0=\mu_{u}(U)=E(L_0+1, U)-E(L_0, U)$]. 

The critical characteristic density $\rho^{l,u}_c(U)$ is, subsequently, obtained via a numerical integration of $n(x)$,
\begin{equation}\label{rho_c}
\rho_c^{l,u}(U)=\frac{N^{l,u}_c(U)}{R}=\frac{1}{R}\int_{-\infty}^{\infty}n\left[\mu_{l,u}(U)-\frac{x^2}{R^2}, U\right] dx\,.
\end{equation}
In this paper, we focus on interaction strengths $U>U^{n=1}_c$ for which there is always a Mott insulator at $n=1$.

\begin{figure}[!t]
\begin{center}
\includegraphics[width=0.99\columnwidth]{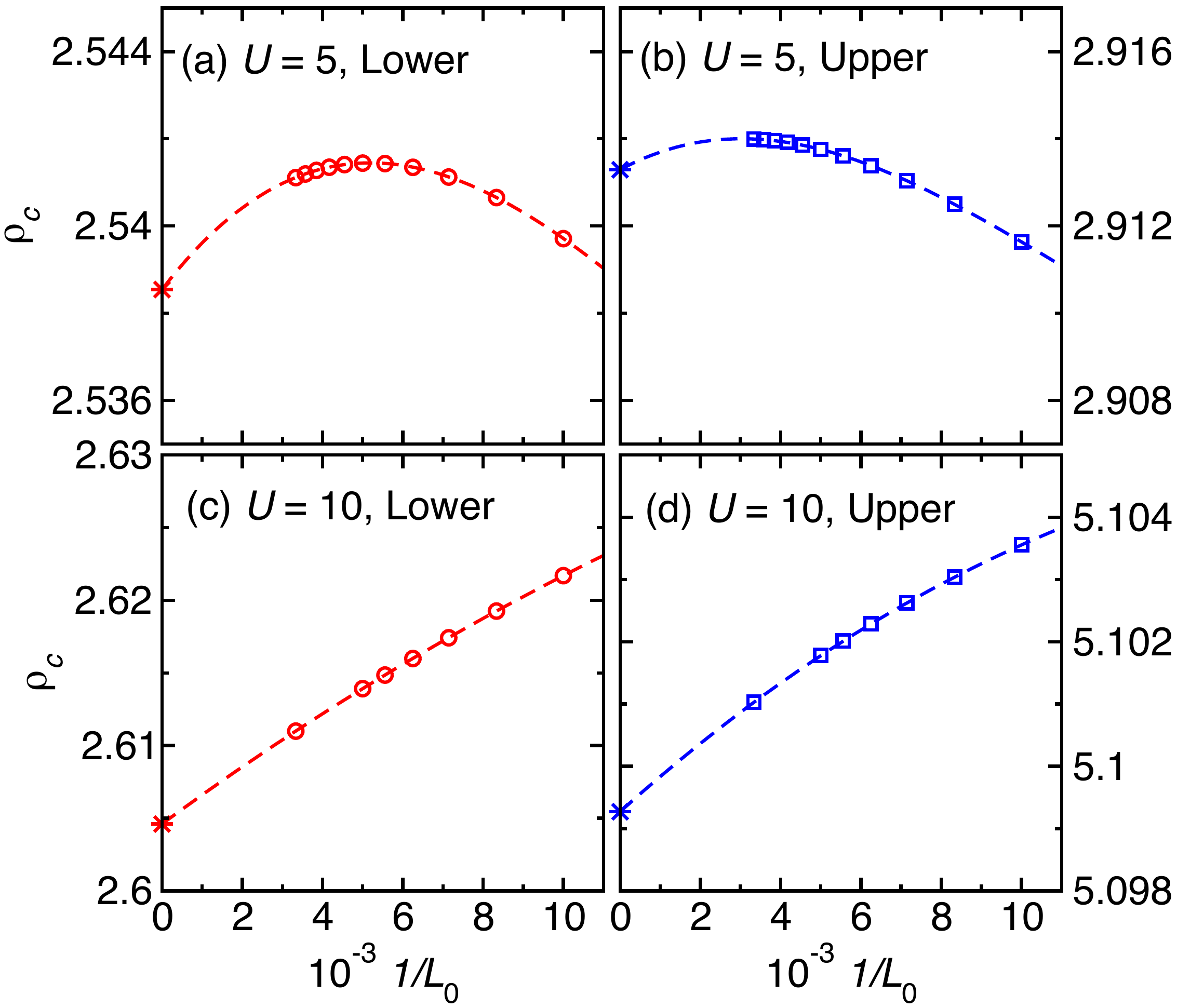}
\caption{Scaling of the LDA results for the critical characteristic densities of the local quantum phase transitions in the trapped Bose-Hubbard model. The open symbols denote results for finite systems with $L_0$ sites, the dashed lines are fits of $\rho^{l,u}_c$ to cubic polynomials in $1/L_0$, and the stars denote the extrapolated results at $L_0\to\infty$. (a) and (c) $\rho^l_c$ for the formation of the $n=1$ Mott insulator at the center of the trap (lower transition) for $U=5$ and 10, respectively. (b) and (d) $\rho^u_c$ for the formation of the $n>1$ superfluid domain at the center of the $n=1$ Mott-insulating one (upper transition) for $U=5$ and 10, respectively.} 
\label{lda}
\end{center}
\end{figure}

A state diagram constructed using the LDA as explained before with results from homogeneous systems with $L_0=200$ is shown in Fig.~\ref{statediagram}. The different states on the $\rho\,$--$\,U$ plane for $U>U_c^{n=1}$ correspond to (see site-occupation profiles in the corresponding insets): (I) $n<1$ superfluid profile, (II) $n=1$ Mott insulator at the center of the trap surrounded by $n<1$ superfluid regions, (III) $n>1$ superfluid region at the center of the $n=1$ Mott-insulating domain that is surrounded by $n<1$ superfluid regions. Note that the lower transition, i.e., the transition from (I) to (II), occurs for values of $\rho^l_c$ that are very close to $\rho_c = 8/\pi$~\cite{zhang_18}. The latter is the critical characteristic density at which a band insulator forms at the center of the trap for spinless fermions [or hard-core bosons, which are the $U\to\infty$ limit of Eq.~(\ref{H_BH})].

State diagrams computed using the LDA based on results from homogeneous systems with finite $L_0$ suffer from finite-size effects as the functions $n(\mu,U)$ and $\mu_{l,u}(U)$ obtained that way are not the ones in the thermodynamic limit. Remarkably, for $L_0>100$, finite-size effects are small in the scale of the state diagram in Fig.~\ref{statediagram}. In Fig.~\ref{LDA}, we show $\rho^{l,u}_c$ calculated from different $n(\mu, U)$'s obtained in systems with up to $L_0=300$ for $U=5$ and $10$. Small but non-negligible differences in $\rho^{l,u}_c$ are found for both transitions. We extrapolate the results to $L_0\to\infty$  (stars in Fig.~\ref{LDA}) by fitting $\rho^{l,u}_c(U)$ with cubic polynomials in $1/L_0$ (dashed lines in Fig.~\ref{LDA}). In Fig.~\ref{statediagram}, we compare the extrapolated results (stars) and the results for $L_0=200$ (triangles). The differences are not noticeable. The results for $\rho^{l,u}_c(U)$ after extrapolation are used as the reference transition points in Sec.~\ref{sec3}.

\section{Local quantum phase transition}\label{sec3}

We turn our attention to using quantum entanglement to detect the transitions shown in Fig.~\ref{statediagram}. Even though those transitions occur locally in space, they still exhibit signatures of quantum phase transitions~\cite{sachdevbook}. In particular, a second derivative of the ground-state energy exhibits indications of nonanalytic behavior at the transition points in the thermodynamic limit.

\begin{figure}[t!]
\begin{center}
\includegraphics[width=0.99\columnwidth]{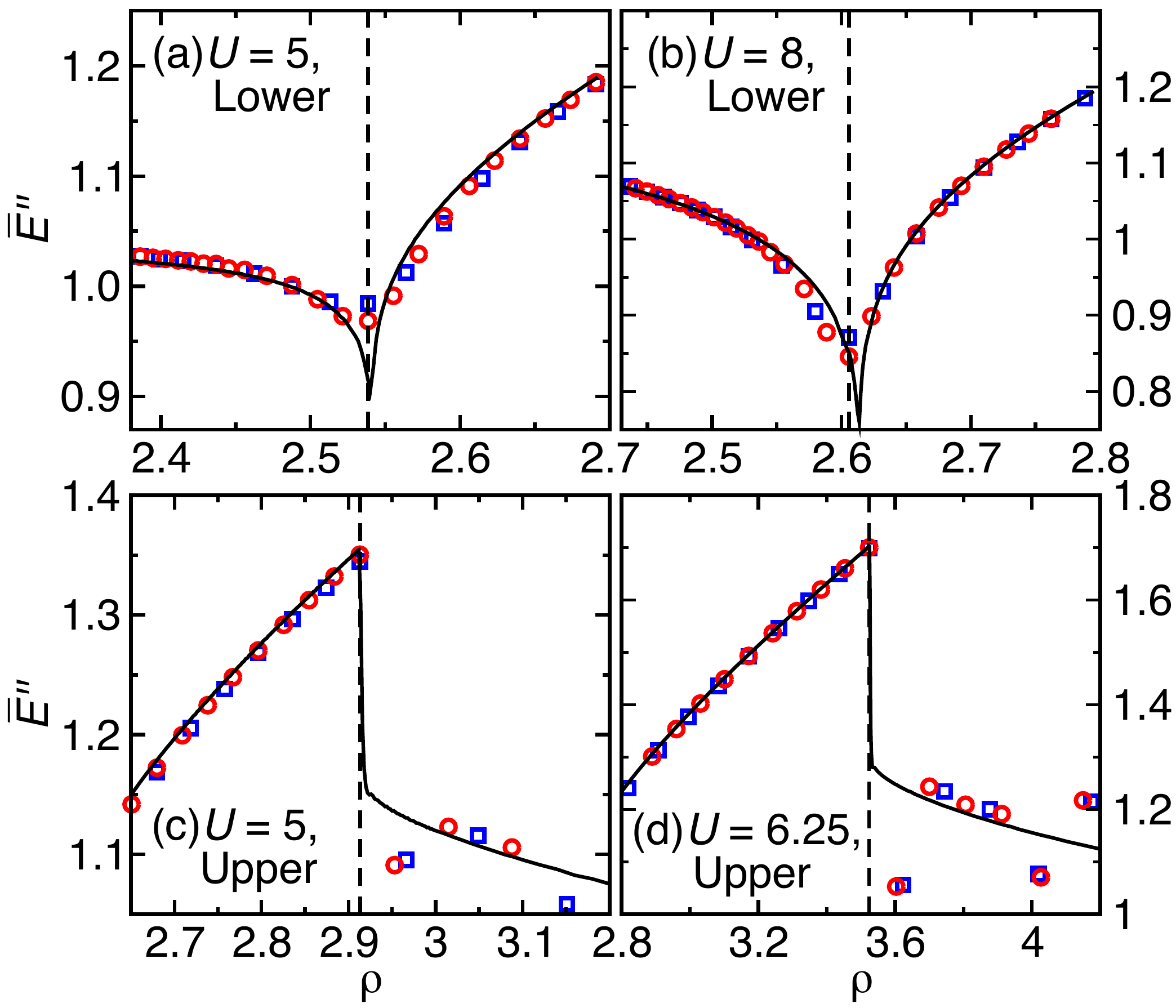}
\caption{Discrete second derivative $\bar E''(\rho)$ of the ground-state energy in the trapped Bose-Hubbard model for (a) $U=5$ and (b) $U=8$ across the lower transition (i.e., the transition in which the $n=1$ Mott insulator forms at the trap center) and for (c) $U=5$ and (d) $U=6.25$ across the upper transition (i.e., the transition in which the $n>1$ superfluid forms within the $n=1$ Mott insulator). To compute the derivative, we fix $R$ and change $N$ in steps of variable numbers of particles to achieve ``closed shells'' in the site occupation profiles. Results are shown for two values of $R$ for each value of $U$ such that at $\rho^{l,u}_c(U)$ (from LDA, vertical dashed lines) the number of particles in the system is $N_c^{l,u}$ as follows. Squares: (a) $N_c^{l}=200$ ($R=78.79$), (b) $N_c^{l}=200$ ($R=76.75$), (c) $N_c^{u}=150$ ($R=51.49$), and (d) $N_c^{u}=80$ ($R=22.70$). Circles: (a) $N_c^{l}=300$ ($R=118.18$), (b) $N_c^{l}=300$ ($R=115.13$), (c) $N_c^{u}=200$ ($R=68.65$), and (d) $N_c^{u}=100$ ($R=28.38$). The solid lines depict results obtained within the LDA (with $L_0=200$). Small finite-size effects are apparent in the departure of the LDA kinks in (a) and (b) from the thermodynamic limit predictions for $\rho^{l}_c$ (dashed line).} 
\label{energyspectrum}
\end{center}
\end{figure}

In Fig.~\ref{energyspectrum}, we show results for the discrete second derivative of the ground-state energy density $\bar E = E/R$ in the presence of the harmonic trap [Hamiltonian~(\ref{H_BH})],
\begin{equation}\label{dE}
\bar E''(\rho) = \frac{E(\rho+\delta\rho)-2E(\rho)+E(\rho-\delta\rho)}{R\; \delta\rho^2}\,,
\end{equation}
across the lower [Figs.~\ref{energyspectrum}(a) and~\ref{energyspectrum}(b)] and upper [Figs.~\ref{energyspectrum}(c) and~\ref{energyspectrum}(d)] transitions. In our calculations, we fix $R$ and change $\rho$ by increasing $N$. The numerical results for $\bar E''(\rho)$ (symbols in Fig.~\ref{energyspectrum}) provide strong indications that, in the thermodynamic limit, $\bar E''(\rho)$ is continuous with a kink developing at $\rho^l_c$ [Figs.~\ref{energyspectrum}(a) and~\ref{energyspectrum}(b)] whereas it is discontinuous at $\rho^u_c$ [Figs.~\ref{energyspectrum}(c) and~\ref{energyspectrum}(d)]. These features are better seen in the results obtained for $\bar E''(\rho)$ within the LDA (reported as continuous lines). The behavior of $\bar E''(\rho)$ at the lower transition [Figs.~\ref{energyspectrum}(a) and~\ref{energyspectrum}(b)] is qualitatively similar to the one found at the critical characteristic density at which the band-insulating domain forms for trapped spinless fermions~\cite{zhang_18}.

The goal of our paper is to use entanglement measures to detect the local quantum phase transitions in the Bose-Hubbard model in the presence of a harmonic trap [see Eq.~\eqref{H_BH}]. We split the system into two halves, $A$ and $\bar A$. For the ground state $|m\rangle$, the reduced density matrix of subsystem $A$ is $\hat\rho_A = \Tr_{\bar A}|m\rangle\langle m|$. We are interested in the von Neumann entanglement entropy,
\begin{equation}\label{svn}
S_{\rm vN}=-\Tr\{ \hat\rho_A\ln\hat\rho_A \} =-\sum_{j}\lambda_{j}\ln\lambda_{j}\,,
\end{equation}
and the more general Renyi entanglement entropy of order $\alpha$ ($S_{\rm vN}$ is the $\alpha\to1$ limit of $S_\alpha$),
\begin{equation}\label{s2}
S_{\alpha}=\frac{1}{1-\alpha}\ln\left[\Tr\{\hat\rho_A^\alpha \}\right] = \frac{1}{1-\alpha}\ln \sum_{j}\lambda_{j}^\alpha\,.
\end{equation}
where $\lambda_{j}$ are the eigenvalues of $\hat \rho_A$, which are computed using DMRG. $S_{2}$ is the Renyi entanglement entropy that was measured in recent quantum gases experiments. 

\begin{figure}[t]
\begin{center}
\includegraphics[width=0.99\columnwidth]{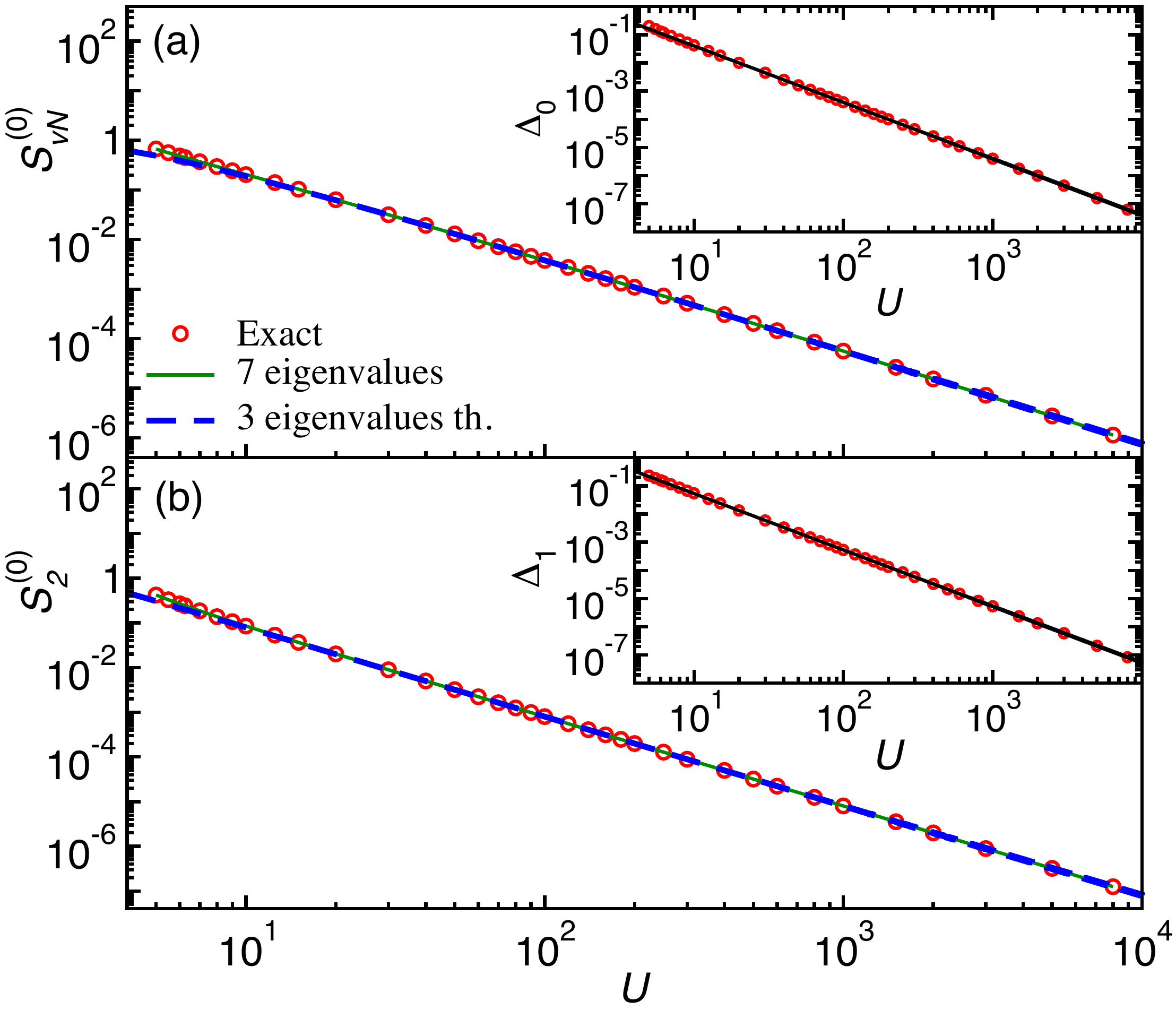}
\caption{Entanglement entropy for the ground state of the homogeneous Bose-Hubbard model [$R^{-2}=0$ in Eq.~(\ref{H_BH})] in the $n=1$ Mott-insulating phase. Main panels: (a) von Neumann entanglement entropy $S^{(0)}_{\rm vN}$ and (b) second-order Renyi entropy $S^{(0)}_{2}$ as functions of the interaction strength $U$. The symbols show the numerical results. The solid lines are results obtained evaluating Eqs.~\eqref{svn} and~\eqref{s2} using only the seven largest eigenvalues of the entanglement spectrum (obtained numerically). The dashed lines are the theoretical results obtained evaluating Eqs.~\eqref{svn} and~\eqref{s2} using only the three largest eigenvalues of the entanglement spectrum with $\Delta_i$ [see Eq.~\eqref{eigenspectrum}] computed to leading order in $1/U$~\cite{alba_12}. The insets: (a) $\Delta_0$ and (b) $\Delta_1$ vs $U$. The symbols depict the numerical results, whereas the solid lines show results of fits to $\Delta_i=\alpha_i/U^2$ for $U\geq100$, which yield $\alpha_0=4.00$ and $\alpha_1=5.33$. The DMRG results were obtained in an equal bipartition of a homogeneous chain with $L_0=200$.} 
\label{homogeneous}
\end{center}
\end{figure}

For trapped spinless fermions in one-dimensional lattices, the formation of a band insulating domain at the center of the trap leads to a vanishing entanglement entropy~\cite{zhang_18}. Similarly, here we expect the formation of a (gapped) Mott-insulating domain with $n=1$ (lower transition) at the center of the trap to result in a reduction of the entanglement entropy. In contrast, we expect the formation of the (critical) superfluid domain with $n>1$ at the center of the Mott-insulating one (upper transition) to result in an increase in the entanglement entropy. We show in Secs.~\ref{sec3_1} and~\ref{sec3_2} that, as a result of the aforementioned expected changes, the entanglement entropy can be used as an order parameter for both local quantum phase transitions. 

We note that when a Mott-insulating domain forms at the center of the trap, due to the presence of a finite correlation length, the bipartite entanglement entropies in the trapped system are expected to approach the values in the Mott-insulating phase in a homogeneous system, which we denote as $S^{(0)}_{\rm vN}$ and $S^{(0)}_{\alpha}$. In the main panel of Fig.~\ref{homogeneous}(a) [Fig.~\ref{homogeneous}(b)], we plot the ground-state von Neumann entanglement entropy $S^{(0)}_{\rm vN}$ (the second Renyi entanglement entropy $S^{(0)}_2$) in the Mott-insulating phase of the homogeneous Bose-Hubbard model vs $U$. $S^{(0)}_{\rm vN}$ and $S^{(0)}_2$ can be seen to vanish as a power law in $1/U$ with the leading-order scaling
\begin{equation}
S^{(0)}_{\rm vN}\sim(\ln U)/U^2\,, \; \; S^{(0)}_2\sim 1/U^2 \,.
\end{equation}

Those scalings can be obtained analytically from a perturbative expansion (in $1/U$) of the entanglement spectrum~\cite{alba_12}. The three lowest eigenvalues have the form
\begin{align}\label{eigenspectrum}
\lambda_0=e^{-\Delta_0}\,,\;\; \lambda_1=\lambda_2=\frac{2}{U^{2}}e^{\Delta_1}\,,
\end{align}
where, to leading order, $\Delta_i$'s ($i=0,1$) scale as $\alpha_i/U^2$. We find, numerically (see the insets in Fig.~\ref{homogeneous}) that $\alpha_0=4.00$ and $\alpha_1=5.33$, which agree with the values $\alpha_0=4$ and $\alpha_1=16/3$ obtained perturbatively~\cite{alba_comm}. 

In the main panels in Figs.~\ref{homogeneous}(a) and~\ref{homogeneous}(b), we compare the numerical results obtained for $S^{(0)}_{\rm vN}$ or $S^{(0)}_{\alpha}$ with results obtained evaluating Eqs.~\eqref{svn} and~\eqref{s2} including only the seven largest eigenvalues (obtained numerically) and the three largest eigenvalues obtained analytically with $\Delta_i=\alpha_i/U^2$ [see Eq.~\eqref{eigenspectrum}]. The agreement between all the results is excellent for $U>10$.

\subsection{Formation of the $n=1$ Mott insulator at the center of the trap}\label{sec3_1}

Here, we study the entanglement entropy across the transition from states (I) to (II) in Fig.~\ref{statediagram}. Namely, the transition in which the $n=1$ Mott-insulating domain forms at the center of the trap (lower transition).

\begin{figure}[!t]
\begin{center}
\includegraphics[width=0.99\columnwidth]{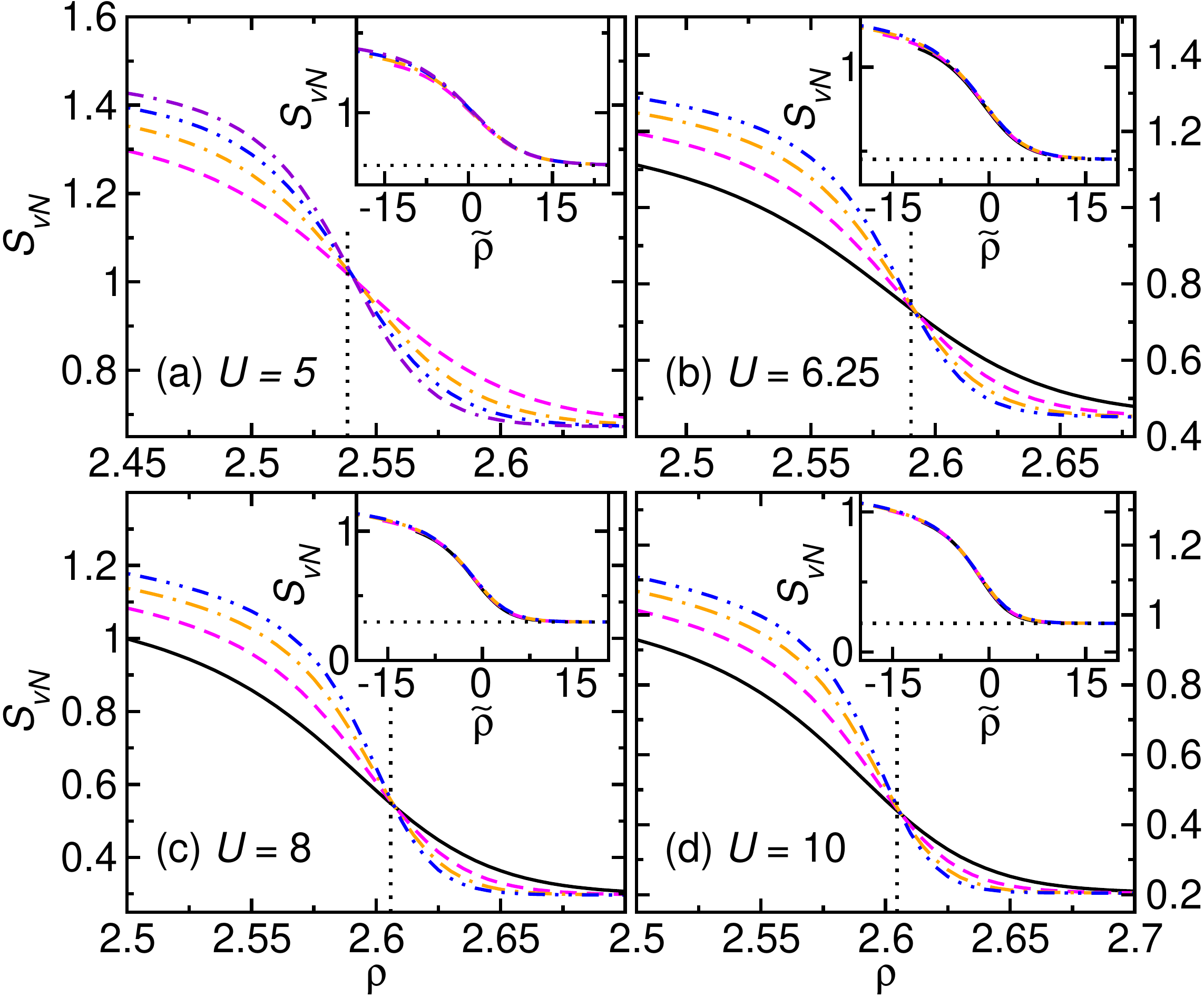}
\caption{$S_{\rm vN}$ for the trapped Bose-Hubbard model when the $n=1$ Mott-insulating domain forms at the center of the trap (lower transition). Main panels: (a) $U=5$ for trapped systems with $N=150$ (dashed line), $N=200$ (dashed-dotted line), $N=250$ (dashed-double-dotted line), and $N=300$ (dotted-double-dashed line) particles. (b)--(d) $U=6.25$, 8, and 10, respectively, for trapped systems with $N=100$ (solid lines), $N=150$ (dashed lines),  $N=200$ (dashed-dotted lines), and $N=250$ (dashed-double-dotted lines) particles. The vertical dotted lines show the LDA predictions for $\rho^l_c$.  The insets: $S_{\rm vN}$ vs $\tilde\rho=(\rho-\rho^l_c)N$ for the same results shown in the corresponding main panels. The horizontal dotted lines show $S_{\rm vN}^{(0)}$ for $n=1$ homogeneous systems with the same values of $U$ [see Fig.~\ref{homogeneous}(a)].} 
\label{lowertransition}
\end{center}
\end{figure}

The main panels of Fig.~\ref{lowertransition} show plots of $S_{\rm vN}$ vs $\rho$ across that transition for four values of the on-site interaction strength $U$. To change $\rho$, we fix the number of trapped particles $N$ and change $R$. In each panel in Fig.~\ref{lowertransition}, we show results for four different values of $N$. For all values of $U$ and $N$, one can see the expected decrease in $S_{\rm vN}$ towards $S^{(0)}_{\rm vN}$ as $\rho$ increases beyond $\rho^l_c$ (predicted by the LDA, vertical dotted lines). We also note that, for each value of $U$, the results for different numbers of particles can be seen to cross very close to the transition point $\rho^l_c$.

Motivated by that crossing, we study the scaling of $S_{\rm vN}$ vs $\rho$ close to $\rho^l_c$. The insets in Fig.~\ref{lowertransition} show plots of $S_{\rm vN}$ vs $\tilde\rho=(\rho-\rho^l_c)N$ that exhibit excellent data collapse. (The collapse improves as $U$ increases as this reduces finite-size effects.) This suggests the existence of a universal scaling function,
\begin{equation}\label{rescaling}
S_{\rm vN}(U)={\cal F}^{(U)}([\rho-\rho^l_c(U)]N)\,
\end{equation} 
for the local quantum phase transition for all interaction strengths $U>U_c$. Note that in the insets, for large values of $\tilde\rho$, $S_{\rm vN}$ converges to $S^{(0)}_{\rm vN}$ (dotted lines) as advanced. As in the study of trapped spinless fermions in Ref.~\cite{zhang_18}, we find that the behavior of the second Renyi entanglement entropy $S_2$ in trapped bosonic systems is qualitatively similar to the one discussed before for $S_{\rm vN}$, so we do not report results for $S_2$ here.

Next, we explore whether measurements of the entanglement entropy in smaller systems, such as the ones that are currently accessible experimentally~\cite{islam_ma_15}, allows one to determine $\rho^l_c$. In order to address this question, we focus on the second Renyi entanglement entropy as this is the one that is of relevance to experiments.

\begin{figure}[!t]
\begin{center}
\includegraphics[width=0.99\columnwidth]{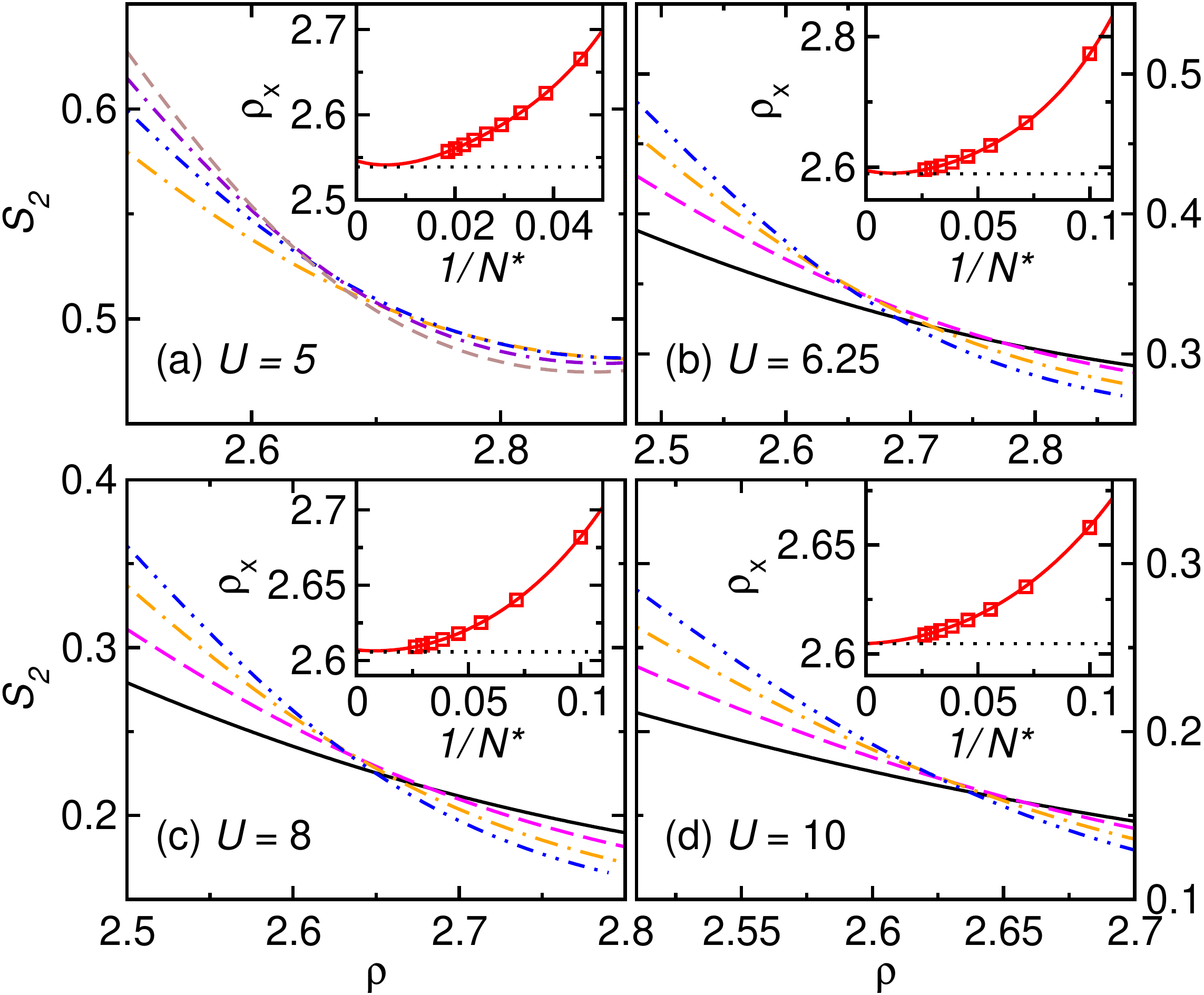}
\caption{$S_2$ vs $\rho$ across the lower transition for small numbers of particles. Main panels: (a) $U=5$ for trapped systems with $N=16$ (dashed-dotted line), $N=20$ (dashed-double-dotted line), $N=24$ (dotted-double-dashed line), and $N=28$ (dashed line) particles. (b)--(d): $U=6.25$, 8, and 10, respectively, for trapped systems with $N=8$ (solid lines), $N=12$ (dashed lines), $N=16$ (dashed-dotted lines), and $N=20$ (dashed-double-dotted lines) particles. The insets: $\rho$ at the crossing points $\rho_{\times}$ of curves for particle numbers $N$ and $N+4$. The values of $U$ are the same as those in the main panels. The solid lines show results of fits of $\rho_{\times}$ to a quartic order polynomial of $1/N^*$ ($N^*=N+2$). The smallest numbers of particles used in the fits are $N=20$ in (a) and $N=8$ in (b)--(d). The horizontal dotted lines are the LDA predictions for $\rho^l_c$.}
\label{smallsystem}
\end{center}
\end{figure}

In Fig.~\ref{smallsystem}, we show results for $S_2$ vs $\rho$ for the same values of $U$ as in Fig.~\ref{lowertransition} but for trapped systems with an order of magnitude smaller number of particles (same order of magnitude but still larger number of particles than in experiments~\cite{islam_ma_15}). As expected because of large finite-size effects, in Fig.~\ref{smallsystem}, the curves for $S_2$ do not cross at the same value of $\rho$ for different values of $N$. The crossing points between curves for the closest number of particles can be seen to move towards smaller values of $\rho$ (towards $\rho^l_c$) as the number of particles increases. 

In the insets in Fig.~\ref{smallsystem}, we plot $\rho$ at the crossing points between curves for systems with $N_1=N$ and $N_2=N+4$ particles (denoted as $\rho_\times$) vs the (inverse) average number of particles $N^*=(N_1+N_2)/2$. We extrapolate the results to $N^*\rightarrow\infty$ by fitting $\rho_{\times}$ vs $1/N^*$ to a quartic order polynomial in $1/N^*$ (dashed lines). The extrapolated results can be seen to give a good estimate of $\rho^l_c$ (shown as horizontal dotted lines), which suggests that experimental measurements of $S_2$ in slightly larger system sizes than those currently accessible could be used to determine $\rho^l_c$ in the future.

The results in Fig.~\ref{smallsystem} make apparent that finite-size effects decrease significantly as $U$ increases and departs from the critical value of $U_c^{n=1}$. As $U$ approaches $U_c^{n=1}$ from above, the insets in Fig.~\ref{smallsystem} show that, for the same numbers of particles, the crossing points $\rho_{\rm x}$ depart from $\rho^l_c$. In addition, for small values of $U$, the curves for the closest number of particles may fail to cross or may not show a clear crossing point because of overlapping for a range of values of $\rho$. For $U=5$, the latter is the case for the $S_2$ curves corresponding to $N=16$ and 20 in Fig.~\ref{smallsystem}(a). This occurs because the lower transition (the formation of the $n=1$ Mott domain) is not well separated from the upper one (the formation of the $n>1$ superfluid domain at the center of the $n=1$ Mott-insulating one) due to the smallness of the Mott gap. As one approaches $U_c^{n=1}$, larger system sizes are needed both experimentally and theoretically to determine $\rho^l_c(U)$.

\subsection{Formation of the $n>1$ superfluid domain at the center of the $n=1$ Mott insulator}\label{sec3_2}

Here, we study the entanglement entropy across the transition from states (II) to (III) in Fig.~\ref{statediagram}. Namely, the transition in which the $n>1$ superfluid domain forms at the center of the $n=1$ Mott-insulating one (upper transition).

\begin{figure}[!t]
\begin{center}
\includegraphics[width=0.99\columnwidth]{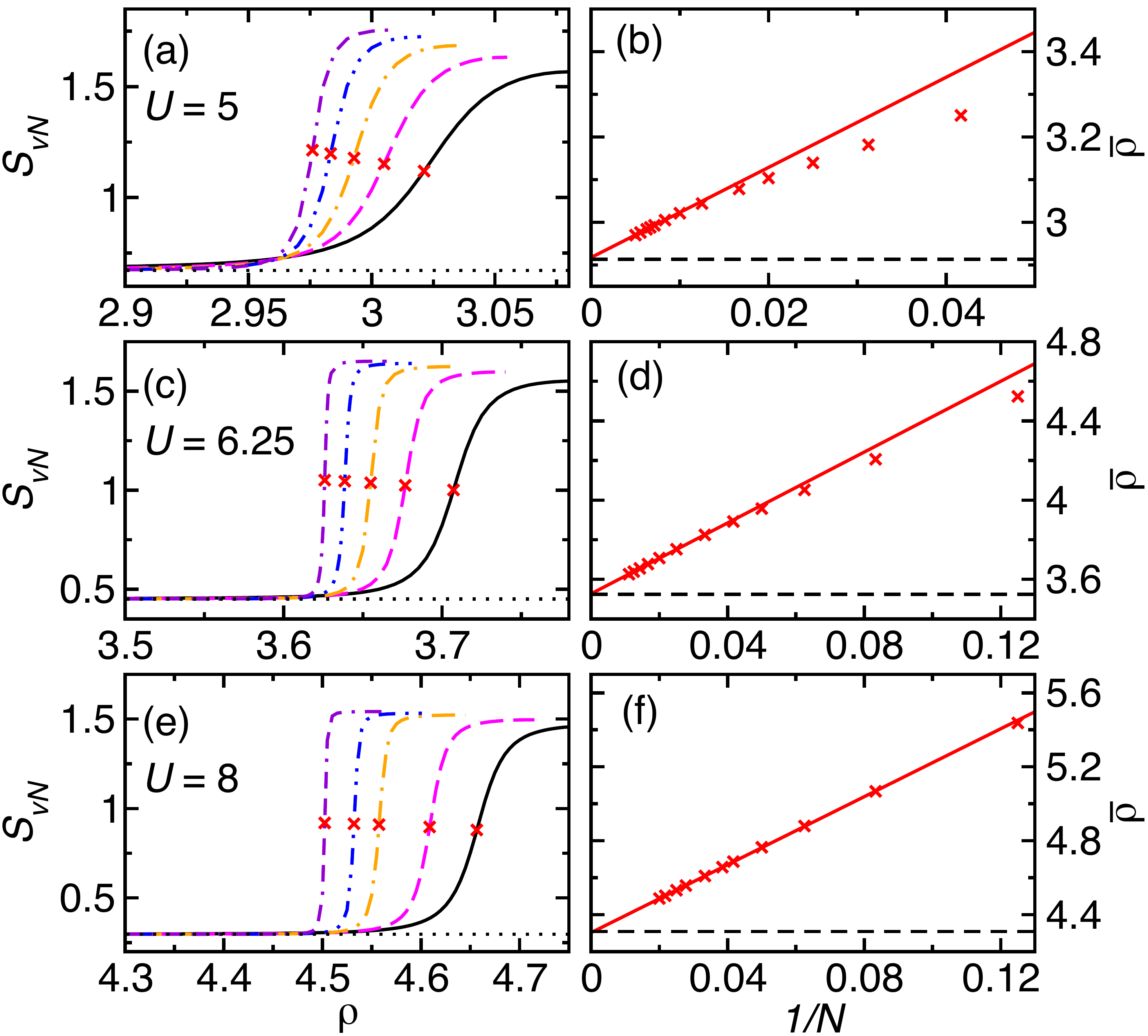}
\caption{(a), (c), (e) $S_{\rm vN}$ in the trapped Bose-Hubbard model across the formation of the $n>1$ superfluid domain at the center of the trap (upper transition). (a) $U=5$ for trapped systems with (from right to left) $N=100$ (solid line), 120 (dashed line), 140 (dashed-dotted line), 160 (dashed-double-dotted line), and 180 (dotted-double-dashed line) particles. (c) $U=6.25$ for trapped systems with (from right to left) $N=50$ (solid line), 60 (dashed line), 70 (dashed-dotted line), 80 (dashed-double-dotted line), and 90 (dotted-double-dashed line) particles. (e) $U=8$ for trapped systems with (from right to left) $N=26$ (solid line), 30 (dashed line), 36 (dashed-dotted line), 40 (dashed-double-dotted line), and 46 (dotted-double-dashed line) particles. The crosses indicate the reference transition points $\bar\rho$ in Eq.~(\ref{refpoint}), and the horizontal dotted lines show $S_{\rm vN}^{(0)}$ for $n=1$ homogeneous systems with the same values of $U$ [see Fig.~\ref{homogeneous}(a)]. For clarity, we cut the $S_{\rm vN}$ vs $\rho$ curves at the maximum values in the plateau. (b), (d), (f) $\bar\rho$ vs $1/N$ for $U=5$, 6.25, and 8, respectively. The solid lines are linear fits to $1/N$ for the six largest values of $N$ shown for each value of $U$. The horizontal dashed lines show the LDA predictions for $\rho^u_c$.} 
\label{uppertransition}
\end{center}
\end{figure}

Figures~\ref{uppertransition}(a),~\ref{uppertransition}(c), and~\ref{uppertransition}(e) show the bipartite entanglement entropy $S_{\rm vN}$ as a function of $\rho$ across the upper transition for $U=5$, 6.25, and 8, respectively. To change $\rho$, we fix the number of trapped particles $N$ and change $R$. In the finite systems studied, the values of $S_{\rm vN}$ can be seen to increase from $S_{\rm vN}^{(0)}$ (horizontal dotted lines) as the site occupations in the center of the trap become larger than one, and then plateau at a value of $S_{\rm vN}$ that depends on the number of particles in the trap. The increase in $S_{\rm vN}$ due to the emergence of the $n>1$ superfluid domain at the center of the trap becomes sharper as $U$ increases and, for each value of $U$, it becomes sharper as $N$ increases. Also, for each value of $U$ as one increases $N$, the sharp increase in $S_{\rm vN}$ occurs at a value of $\rho$ that approaches the LDA prediction $\rho_c^u$ in the thermodynamic limit. The sharp increase observed in $S_{\rm vN}$ as $N$ increases for each value of $U$ resembles the jump of $\bar E''(\rho)$ seen in Figs.~\ref{energyspectrum}(c) and~\ref{energyspectrum}(d) at $\rho_c^u$.

In what follows, we directly use the sharp increase in $S_{\rm vN}$ when $n>1$ to carry out scaling analyses to predict the critical characteristic densities $\rho^u_c(U)$ in the thermodynamic limit. We choose reference transition points $\bar \rho$ for each value of $N$, see the crosses in Figs.~\ref{uppertransition}(a),~\ref{uppertransition}(c), and~\ref{uppertransition}(e), defined as
\begin{equation}\label{refpoint}
S_{\rm vN}(\bar \rho)=(S_{\rm vN}^{(0)}+S^{\rm max}_{\rm vN})/2\,,
\end{equation}
where $S_{\rm vN}^{(0)}$ ($S^{\rm max}$) is the minimum (maximum) $S_{\rm vN}$ right before (after) the rapid increase.

In Figs.~\ref{uppertransition}(b),~\ref{uppertransition}(d), and~\ref{uppertransition}(f), we show how $\bar\rho$ changes as one changes the number of particles in the trap for the same values of $U$ as in Figs.~\ref{uppertransition}(a),~\ref{uppertransition}(c), and~\ref{uppertransition}(e), respectively. We observe that $\bar \rho$ moves towards the LDA predicted $\rho^u_c$ (see the horizontal dashed lines) when increasing $N$. A linear fit of $\bar\rho$ vs $1/N$ for largest number of particles reported in the figures is used to extrapolate $\rho^u_c$ to $N\to\infty$ (see the solid lines). The results of the extrapolations agree well with the LDA predictions. This shows that one can use $S_{\rm vN}$ from calculations in finite systems to determine the upper transition for trapped bosons in the thermodynamic limit. Moreover, for large values of $U$ [see Fig.~\ref{uppertransition}(f) for $U=8$], the linear relation between $\bar\rho$ and $1/N$ extends to the system sizes accessible in current experiments ($N\sim 10$). 

We have checked that the second Renyi entanglement entropy (not shown here) exhibits a behavior across the upper transition that is qualitatively similar to the one shown for the von Neumann entanglement entropy as also found across the lower transition.

\section{Summary} \label{sec_conclusion}

We studied bipartite entanglement entropies in the ground state of the one-dimensional Bose-Hubbard model in the presence of a harmonic confining potential. When increasing the characteristic density, this inhomogeneous system undergoes local quantum phase transitions that are characterized by nonanalytic behaviors of the ground-state energy. Specifically, we showed that the second derivative of the total energy with respect to the characteristic density is continuous with a kink at the lower transition and that it is discontinuous at the upper transition. We also showed that bipartite entanglement entropies serve as order parameters for those local quantum phase transitions.

The first transition studied is the formation of the $n=1$ Mott insulator at the center of the trap (lower transition). We showed that this transition is accompanied by a smooth decrease in the entanglement entropy to the value of the corresponding Mott-insulating phase in the homogeneous system. A simple rescaling of the characteristic density resulted in data collapse for the various entanglement entropy curves for different system sizes about the critical characteristic density $\rho_c^l$. The second transition studied is the emergence of the $n>1$ superfluid domain at the center of the Mott-insulating one (upper transition). In this case, the entanglement entropy exhibits a sharp increase at a characteristic density that approaches the critical characteristic density $\rho_c^u$ as one increases the number of particles in the trap. We used an extrapolation scheme to determine the critical characteristic density $\rho_c^u$ from the finite-system-size calculations that yielded accurate results.

A special focus was devoted to entanglement entropies of small systems, of relevance to current ultracold gases experiments~\cite{islam_ma_15, kaufman_tai_16}. We showed that, for sufficiently large values of $U$,  the critical characteristic densities for both transitions in the thermodynamic limit can be predicted using appropriate finite-size scaling analyses of the entanglement entropies. Despite the fact that we only studied ground-state properties, we expect the observed behavior of the entanglement entropies across the transitions to be preserved at low temperatures~\cite{zhang_18}, which is expected to be the case in experiments.

\section{Acknowledgments}
We thank V. Alba, F. Heidrich-Meisner, and R. Modak for discussions. Y.Z. and M.R. acknowledge support from NSF Grant No.~PHY-1707482. L.V. acknowledges support from the Slovenian Research Agency (ARRS), research core fundings Grants No.~P1-0044 and No.~J1-1696.




\bibliographystyle{biblev1}
\bibliography{references}

\begin{thebibliography}{10}
\expandafter\ifx\csname url\endcsname\relax
  \def\url#1{{\tt #1}}\fi
\expandafter\ifx\csname urlprefix\endcsname\relax\def\urlprefix{URL }\fi
\expandafter\ifx\csname bibinfo\endcsname\relax\def\bibinfo#1#2{#2}\fi
\expandafter\ifx\csname eprint\endcsname\relax\def\eprint#1{\url{#1}}\fi

\bibitem{horodecki_09}
\bibinfo{author}{R.~Horodecki}, \bibinfo{author}{P.~Horodecki},
  \bibinfo{author}{M.~Horodecki}, and \bibinfo{author}{K.~Horodecki},
  \bibinfo{title}{Quantum entanglement},
  \bibinfo{journal}{\href{http://dx.doi.org/10.1103/RevModPhys.81.865}{Rev.
  Mod. Phys.}} \href{http://dx.doi.org/10.1103/RevModPhys.81.865}{{\bf
  \bibinfo{volume}{81}}, \bibinfo{pages}{865}}
  (\href{http://dx.doi.org/10.1103/RevModPhys.81.865}{\bibinfo{year}{2009}}).

\bibitem{amico_fazio_08}
\bibinfo{author}{L.~Amico}, \bibinfo{author}{R.~Fazio},
  \bibinfo{author}{A.~Osterloh}, and \bibinfo{author}{V.~Vedral},
  \bibinfo{title}{Entanglement in many-body systems},
  \bibinfo{journal}{\href{http://dx.doi.org/10.1103/RevModPhys.80.517}{Rev.
  Mod. Phys.}} \href{http://dx.doi.org/10.1103/RevModPhys.80.517}{{\bf
  \bibinfo{volume}{80}}, \bibinfo{pages}{517}}
  (\href{http://dx.doi.org/10.1103/RevModPhys.80.517}{\bibinfo{year}{2008}}).

\bibitem{eisert_cramer_10}
\bibinfo{author}{J.~Eisert}, \bibinfo{author}{M.~Cramer}, and
  \bibinfo{author}{M.~B. Plenio}, \bibinfo{title}{Colloquium: Area laws for the
  entanglement entropy},
  \bibinfo{journal}{\href{http://dx.doi.org/10.1103/RevModPhys.82.277}{Rev.
  Mod. Phys.}} \href{http://dx.doi.org/10.1103/RevModPhys.82.277}{{\bf
  \bibinfo{volume}{82}}, \bibinfo{pages}{277}}
  (\href{http://dx.doi.org/10.1103/RevModPhys.82.277}{\bibinfo{year}{2010}}).

\bibitem{laflorencie_16}
\bibinfo{author}{N.~Laflorencie}, \bibinfo{title}{Quantum entanglement in
  condensed matter systems},
  \bibinfo{journal}{\href{http://dx.doi.org/10.1016/j.physrep.2016.06.008}{Physics
  Reports}} \href{http://dx.doi.org/10.1016/j.physrep.2016.06.008}{{\bf
  \bibinfo{volume}{646}}, \bibinfo{pages}{1 }}
  (\href{http://dx.doi.org/10.1016/j.physrep.2016.06.008}{\bibinfo{year}{2016}}).

\bibitem{osterloh_amico_2002}
\bibinfo{author}{A.~Osterloh}, \bibinfo{author}{L.~Amico},
  \bibinfo{author}{G.~Falci}, and \bibinfo{author}{R.~Fazio},
  \bibinfo{title}{Scaling of entanglement close to a quantum phase transition},
  \bibinfo{journal}{\href{http://dx.doi.org/10.1038/416608a}{Nature}}
  \href{http://dx.doi.org/10.1038/416608a}{{\bf \bibinfo{volume}{416}},
  \bibinfo{pages}{608}}
  (\href{http://dx.doi.org/10.1038/416608a}{\bibinfo{year}{2002}}).

\bibitem{osborne_nielsen_02}
\bibinfo{author}{T.~J. Osborne} and \bibinfo{author}{M.~A. Nielsen},
  \bibinfo{title}{Entanglement in a simple quantum phase transition},
  \bibinfo{journal}{\href{http://dx.doi.org/10.1103/PhysRevA.66.032110}{Phys.
  Rev. A}} \href{http://dx.doi.org/10.1103/PhysRevA.66.032110}{{\bf
  \bibinfo{volume}{66}}, \bibinfo{pages}{032110}}
  (\href{http://dx.doi.org/10.1103/PhysRevA.66.032110}{\bibinfo{year}{2002}}).

\bibitem{vidal_03}
\bibinfo{author}{G.~Vidal}, \bibinfo{author}{J.~I. Latorre},
  \bibinfo{author}{E.~Rico}, and \bibinfo{author}{A.~Kitaev},
  \bibinfo{title}{Entanglement in quantum critical phenomena},
  \bibinfo{journal}{\href{http://dx.doi.org/10.1103/PhysRevLett.90.227902}{Phys.
  Rev. Lett.}} \href{http://dx.doi.org/10.1103/PhysRevLett.90.227902}{{\bf
  \bibinfo{volume}{90}}, \bibinfo{pages}{227902}}
  (\href{http://dx.doi.org/10.1103/PhysRevLett.90.227902}{\bibinfo{year}{2003}}).

\bibitem{calabrese_04}
\bibinfo{author}{P.~Calabrese} and \bibinfo{author}{J.~Cardy},
  \bibinfo{title}{Entanglement entropy and quantum field theory},
  \bibinfo{journal}{\href{http://dx.doi.org/10.1088/1742-5468/2004/06/p06002}{J.
  Stat. Mech.}} \href{http://dx.doi.org/10.1088/1742-5468/2004/06/p06002}{{\bf
  \bibinfo{volume}{{\rm (2004)}}}, \bibinfo{pages}{P06002}}.

\bibitem{calabrese_2005}
\bibinfo{author}{P.~Calabrese} and \bibinfo{author}{J.~Cardy},
  \bibinfo{title}{Evolution of entanglement entropy in one-dimensional
  systems},
  \bibinfo{journal}{\href{http://dx.doi.org/10.1088/1742-5468/2005/04/p04010}{J.
  Stat. Mech.}} \href{http://dx.doi.org/10.1088/1742-5468/2005/04/p04010}{{\bf
  \bibinfo{volume}{{\rm (2005)}}}, \bibinfo{pages}{P04010}}.

\bibitem{chiara_2006}
\bibinfo{author}{G.~D. Chiara}, \bibinfo{author}{S.~Montangero},
  \bibinfo{author}{P.~Calabrese}, and \bibinfo{author}{R.~Fazio},
  \bibinfo{title}{Entanglement entropy dynamics of {Heisenberg} chains},
  \bibinfo{journal}{\href{http://dx.doi.org/10.1088/1742-5468/2006/03/p03001}{J.
  Stat. Mech.}} \href{http://dx.doi.org/10.1088/1742-5468/2006/03/p03001}{{\bf
  \bibinfo{volume}{{\rm (2006)}}}, \bibinfo{pages}{P03001}}.

\bibitem{Daley_2012}
\bibinfo{author}{A.~J. Daley}, \bibinfo{author}{H.~Pichler},
  \bibinfo{author}{J.~Schachenmayer}, and \bibinfo{author}{P.~Zoller},
  \bibinfo{title}{{Measuring Entanglement Growth in Quench Dynamics of Bosons
  in an Optical Lattice}},
  \bibinfo{journal}{\href{http://dx.doi.org/10.1103/PhysRevLett.109.020505}{Phys.
  Rev. Lett.}} \href{http://dx.doi.org/10.1103/PhysRevLett.109.020505}{{\bf
  \bibinfo{volume}{109}}, \bibinfo{pages}{020505}}
  (\href{http://dx.doi.org/10.1103/PhysRevLett.109.020505}{\bibinfo{year}{2012}}).

\bibitem{Schachenmayer_2013}
\bibinfo{author}{J.~Schachenmayer}, \bibinfo{author}{B.~P. Lanyon},
  \bibinfo{author}{C.~F. Roos}, and \bibinfo{author}{A.~J. Daley},
  \bibinfo{title}{{Entanglement Growth in Quench Dynamics with Variable Range
  Interactions}},
  \bibinfo{journal}{\href{http://dx.doi.org/10.1103/PhysRevX.3.031015}{Phys.
  Rev. X}} \href{http://dx.doi.org/10.1103/PhysRevX.3.031015}{{\bf
  \bibinfo{volume}{3}}, \bibinfo{pages}{031015}}
  (\href{http://dx.doi.org/10.1103/PhysRevX.3.031015}{\bibinfo{year}{2013}}).

\bibitem{kim_2013}
\bibinfo{author}{H.~Kim} and \bibinfo{author}{D.~A. Huse},
  \bibinfo{title}{{Ballistic Spreading of Entanglement in a Diffusive
  Nonintegrable System}},
  \bibinfo{journal}{\href{http://dx.doi.org/10.1103/PhysRevLett.111.127205}{Phys.
  Rev. Lett.}} \href{http://dx.doi.org/10.1103/PhysRevLett.111.127205}{{\bf
  \bibinfo{volume}{111}}, \bibinfo{pages}{127205}}
  (\href{http://dx.doi.org/10.1103/PhysRevLett.111.127205}{\bibinfo{year}{2013}}).

\bibitem{hamma_05}
\bibinfo{author}{A.~Hamma}, \bibinfo{author}{R.~Ionicioiu}, and
  \bibinfo{author}{P.~Zanardi}, \bibinfo{title}{{Ground state entanglement and
  geometric entropy in the Kitaev model}},
  \bibinfo{journal}{\href{http://dx.doi.org/10.1016/j.physleta.2005.01.060}{Phys.
  Lett. A}} \href{http://dx.doi.org/10.1016/j.physleta.2005.01.060}{{\bf
  \bibinfo{volume}{337}}, \bibinfo{pages}{22}}
  (\href{http://dx.doi.org/10.1016/j.physleta.2005.01.060}{\bibinfo{year}{2005}}).

\bibitem{kitaev_06}
\bibinfo{author}{A.~Kitaev} and \bibinfo{author}{J.~Preskill},
  \bibinfo{title}{{Topological Entanglement Entropy}},
  \bibinfo{journal}{\href{http://dx.doi.org/10.1103/PhysRevLett.96.110404}{Phys.
  Rev. Lett.}} \href{http://dx.doi.org/10.1103/PhysRevLett.96.110404}{{\bf
  \bibinfo{volume}{96}}, \bibinfo{pages}{110404}}
  (\href{http://dx.doi.org/10.1103/PhysRevLett.96.110404}{\bibinfo{year}{2006}}).

\bibitem{levin_06}
\bibinfo{author}{M.~Levin} and \bibinfo{author}{X.-G. Wen},
  \bibinfo{title}{{Detecting Topological Order in a Ground State Wave
  Function}},
  \bibinfo{journal}{\href{http://dx.doi.org/10.1103/PhysRevLett.96.110405}{Phys.
  Rev. Lett.}} \href{http://dx.doi.org/10.1103/PhysRevLett.96.110405}{{\bf
  \bibinfo{volume}{96}}, \bibinfo{pages}{110405}}
  (\href{http://dx.doi.org/10.1103/PhysRevLett.96.110405}{\bibinfo{year}{2006}}).

\bibitem{islam_ma_15}
\bibinfo{author}{R.~Islam}, \bibinfo{author}{R.~Ma}, \bibinfo{author}{P.~M.
  Preiss}, \bibinfo{author}{M.~E. Tai}, \bibinfo{author}{A.~Lukin},
  \bibinfo{author}{M.~Rispoli}, and \bibinfo{author}{M.~Greiner},
  \bibinfo{title}{Measuring entanglement entropy in a quantum many-body
  system},
  \bibinfo{journal}{\href{http://dx.doi.org/10.1038/nature15750}{Nature}}
  \href{http://dx.doi.org/10.1038/nature15750}{{\bf \bibinfo{volume}{528}},
  \bibinfo{pages}{77}}
  (\href{http://dx.doi.org/10.1038/nature15750}{\bibinfo{year}{2015}}).

\bibitem{kaufman_tai_16}
\bibinfo{author}{A.~M. Kaufman}, \bibinfo{author}{M.~E. Tai},
  \bibinfo{author}{A.~Lukin}, \bibinfo{author}{M.~Rispoli},
  \bibinfo{author}{R.~Schittko}, \bibinfo{author}{P.~M. Preiss}, and
  \bibinfo{author}{M.~Greiner}, \bibinfo{title}{Quantum thermalization through
  entanglement in an isolated many-body system},
  \bibinfo{journal}{\href{http://dx.doi.org/10.1126/science.aaf6725}{Science}}
  \href{http://dx.doi.org/10.1126/science.aaf6725}{{\bf \bibinfo{volume}{353}},
  \bibinfo{pages}{794}}
  (\href{http://dx.doi.org/10.1126/science.aaf6725}{\bibinfo{year}{2016}}).

\bibitem{brydges_19}
\bibinfo{author}{T.~Brydges}, \bibinfo{author}{A.~Elben},
  \bibinfo{author}{P.~Jurcevic}, \bibinfo{author}{B.~Vermersch},
  \bibinfo{author}{C.~Maier}, \bibinfo{author}{B.~P. Lanyon},
  \bibinfo{author}{P.~Zoller}, \bibinfo{author}{R.~Blatt}, and
  \bibinfo{author}{C.~F. Roos}, \bibinfo{title}{{Probing R{\'e}nyi entanglement
  entropy via randomized measurements}},
  \bibinfo{journal}{\href{http://dx.doi.org/10.1126/science.aau4963}{Science}}
  \href{http://dx.doi.org/10.1126/science.aau4963}{{\bf \bibinfo{volume}{364}},
  \bibinfo{pages}{260}}
  (\href{http://dx.doi.org/10.1126/science.aau4963}{\bibinfo{year}{2019}}).

\bibitem{batrouni_rousseau_02}
\bibinfo{author}{G.~G. Batrouni}, \bibinfo{author}{V.~Rousseau},
  \bibinfo{author}{R.~T. Scalettar}, \bibinfo{author}{M.~Rigol},
  \bibinfo{author}{A.~Muramatsu}, \bibinfo{author}{P.~J.~H. Denteneer}, and
  \bibinfo{author}{M.~Troyer}, \bibinfo{title}{{Mott Domains of Bosons Confined
  on Optical Lattices}},
  \bibinfo{journal}{\href{http://dx.doi.org/10.1103/PhysRevLett.89.117203}{Phys.
  Rev. Lett.}} \href{http://dx.doi.org/10.1103/PhysRevLett.89.117203}{{\bf
  \bibinfo{volume}{89}}, \bibinfo{pages}{117203}}
  (\href{http://dx.doi.org/10.1103/PhysRevLett.89.117203}{\bibinfo{year}{2002}}).

\bibitem{kashurnikov_prokofev_02}
\bibinfo{author}{V.~A. Kashurnikov}, \bibinfo{author}{N.~V. Prokof'ev}, and
  \bibinfo{author}{B.~V. Svistunov}, \bibinfo{title}{Revealing the
  superfluid--{M}ott-insulator transition in an optical lattice},
  \bibinfo{journal}{\href{http://dx.doi.org/10.1103/PhysRevA.66.031601}{Phys.
  Rev. A}} \href{http://dx.doi.org/10.1103/PhysRevA.66.031601}{{\bf
  \bibinfo{volume}{66}}, \bibinfo{pages}{031601}}
  (\href{http://dx.doi.org/10.1103/PhysRevA.66.031601}{\bibinfo{year}{2002}}).

\bibitem{rigol_muramatsu_03}
\bibinfo{author}{M.~Rigol}, \bibinfo{author}{A.~Muramatsu},
  \bibinfo{author}{G.~G. Batrouni}, and \bibinfo{author}{R.~T. Scalettar},
  \bibinfo{title}{{Local Quantum Criticality in Confined Fermions on Optical
  Lattices}},
  \bibinfo{journal}{\href{http://dx.doi.org/10.1103/PhysRevLett.91.130403}{Phys.
  Rev. Lett.}} \href{http://dx.doi.org/10.1103/PhysRevLett.91.130403}{{\bf
  \bibinfo{volume}{91}}, \bibinfo{pages}{130403}}
  (\href{http://dx.doi.org/10.1103/PhysRevLett.91.130403}{\bibinfo{year}{2003}}).

\bibitem{rigol_muramatsu_04may}
\bibinfo{author}{M.~Rigol} and \bibinfo{author}{A.~Muramatsu},
  \bibinfo{title}{Quantum {M}onte {C}arlo study of confined fermions in
  one-dimensional optical lattices},
  \bibinfo{journal}{\href{http://dx.doi.org/10.1103/PhysRevA.69.053612}{Phys.
  Rev. A}} \href{http://dx.doi.org/10.1103/PhysRevA.69.053612}{{\bf
  \bibinfo{volume}{69}}, \bibinfo{pages}{053612}}
  (\href{http://dx.doi.org/10.1103/PhysRevA.69.053612}{\bibinfo{year}{2004}}).

\bibitem{kollath_schollwoeck_04}
\bibinfo{author}{C.~Kollath}, \bibinfo{author}{U.~Schollw\"ock},
  \bibinfo{author}{J.~von Delft}, and \bibinfo{author}{W.~Zwerger},
  \bibinfo{title}{Spatial correlations of trapped one-dimensional bosons in an
  optical lattice},
  \bibinfo{journal}{\href{http://dx.doi.org/10.1103/PhysRevA.69.031601}{Phys.
  Rev. A}} \href{http://dx.doi.org/10.1103/PhysRevA.69.031601}{{\bf
  \bibinfo{volume}{69}}, \bibinfo{pages}{031601}}
  (\href{http://dx.doi.org/10.1103/PhysRevA.69.031601}{\bibinfo{year}{2004}}).

\bibitem{rigol_muramatsu_04b}
\bibinfo{author}{M.~Rigol} and \bibinfo{author}{A.~Muramatsu},
  \bibinfo{title}{Confinement control by optical lattices},
  \bibinfo{journal}{\href{http://dx.doi.org/10.1103/PhysRevA.70.043627}{Phys.
  Rev. A}} \href{http://dx.doi.org/10.1103/PhysRevA.70.043627}{{\bf
  \bibinfo{volume}{70}}, \bibinfo{pages}{043627}}
  (\href{http://dx.doi.org/10.1103/PhysRevA.70.043627}{\bibinfo{year}{2004}}).

\bibitem{hooley_quintanilla_04}
\bibinfo{author}{C.~Hooley} and \bibinfo{author}{J.~Quintanilla},
  \bibinfo{title}{Single-atom density of states of an optical lattice},
  \bibinfo{journal}{\href{http://dx.doi.org/10.1103/PhysRevLett.93.080404}{Phys.
  Rev. Lett.}} \href{http://dx.doi.org/10.1103/PhysRevLett.93.080404}{{\bf
  \bibinfo{volume}{93}}, \bibinfo{pages}{080404}}
  (\href{http://dx.doi.org/10.1103/PhysRevLett.93.080404}{\bibinfo{year}{2004}}).

\bibitem{bergkvist_henelius_04}
\bibinfo{author}{S.~Bergkvist}, \bibinfo{author}{P.~Henelius}, and
  \bibinfo{author}{A.~Rosengren}, \bibinfo{title}{Local-density approximation
  for confined bosons in an optical lattice},
  \bibinfo{journal}{\href{http://dx.doi.org/10.1103/PhysRevA.70.053601}{Phys.
  Rev. A}} \href{http://dx.doi.org/10.1103/PhysRevA.70.053601}{{\bf
  \bibinfo{volume}{70}}, \bibinfo{pages}{053601}}
  (\href{http://dx.doi.org/10.1103/PhysRevA.70.053601}{\bibinfo{year}{2004}}).

\bibitem{wessel_alet_04}
\bibinfo{author}{S.~Wessel}, \bibinfo{author}{F.~Alet},
  \bibinfo{author}{M.~Troyer}, and \bibinfo{author}{G.~G. Batrouni},
  \bibinfo{title}{{Quantum Monte Carlo simulations of confined bosonic atoms in
  optical lattices}},
  \bibinfo{journal}{\href{http://dx.doi.org/10.1103/PhysRevA.70.053615}{Phys.
  Rev. A}} \href{http://dx.doi.org/10.1103/PhysRevA.70.053615}{{\bf
  \bibinfo{volume}{70}}, \bibinfo{pages}{053615}}
  (\href{http://dx.doi.org/10.1103/PhysRevA.70.053615}{\bibinfo{year}{2004}}).

\bibitem{rigol_muramatsu_04sept}
\bibinfo{author}{M.~Rigol} and \bibinfo{author}{A.~Muramatsu},
  \bibinfo{title}{Universal properties of hard-core bosons confined on
  one-dimensional lattices},
  \bibinfo{journal}{\href{http://dx.doi.org/10.1103/PhysRevA.70.031603}{Phys.
  Rev. A}} \href{http://dx.doi.org/10.1103/PhysRevA.70.031603}{{\bf
  \bibinfo{volume}{70}}, \bibinfo{pages}{031603}}
  (\href{http://dx.doi.org/10.1103/PhysRevA.70.031603}{\bibinfo{year}{2004}}).

\bibitem{rigol_muramatsu_05july}
\bibinfo{author}{M.~Rigol} and \bibinfo{author}{A.~Muramatsu},
  \bibinfo{title}{Ground-state properties of hard-core bosons confined on
  one-dimensional optical lattices},
  \bibinfo{journal}{\href{http://dx.doi.org/10.1103/PhysRevA.72.013604}{Phys.
  Rev. A}} \href{http://dx.doi.org/10.1103/PhysRevA.72.013604}{{\bf
  \bibinfo{volume}{72}}, \bibinfo{pages}{013604}}
  (\href{http://dx.doi.org/10.1103/PhysRevA.72.013604}{\bibinfo{year}{2005}}).

\bibitem{rey_pupillo_05}
\bibinfo{author}{A.~M. Rey}, \bibinfo{author}{G.~Pupillo},
  \bibinfo{author}{C.~W. Clark}, and \bibinfo{author}{C.~J. Williams},
  \bibinfo{title}{Ultracold atoms confined in an optical lattice plus parabolic
  potential: A closed-form approach},
  \bibinfo{journal}{\href{http://dx.doi.org/10.1103/PhysRevA.72.033616}{Phys.
  Rev. A}} \href{http://dx.doi.org/10.1103/PhysRevA.72.033616}{{\bf
  \bibinfo{volume}{72}}, \bibinfo{pages}{033616}}
  (\href{http://dx.doi.org/10.1103/PhysRevA.72.033616}{\bibinfo{year}{2005}}).

\bibitem{batrouni_krishnamurthy_08}
\bibinfo{author}{G.~G. Batrouni}, \bibinfo{author}{H.~R. Krishnamurthy},
  \bibinfo{author}{K.~W. Mahmud}, \bibinfo{author}{V.~G. Rousseau}, and
  \bibinfo{author}{R.~T. Scalettar}, \bibinfo{title}{Canonical trajectories and
  critical coupling of the {Bose-Hubbard Hamiltonian} in a harmonic trap},
  \bibinfo{journal}{\href{http://dx.doi.org/10.1103/PhysRevA.78.023627}{Phys.
  Rev. A}} \href{http://dx.doi.org/10.1103/PhysRevA.78.023627}{{\bf
  \bibinfo{volume}{78}}, \bibinfo{pages}{023627}}
  (\href{http://dx.doi.org/10.1103/PhysRevA.78.023627}{\bibinfo{year}{2008}}).

\bibitem{rigol_batrouni_09}
\bibinfo{author}{M.~Rigol}, \bibinfo{author}{G.~G. Batrouni},
  \bibinfo{author}{V.~G. Rousseau}, and \bibinfo{author}{R.~T. Scalettar},
  \bibinfo{title}{State diagrams for harmonically trapped bosons in optical
  lattices},
  \bibinfo{journal}{\href{http://dx.doi.org/10.1103/PhysRevA.79.053605}{Phys.
  Rev. A}} \href{http://dx.doi.org/10.1103/PhysRevA.79.053605}{{\bf
  \bibinfo{volume}{79}}, \bibinfo{pages}{053605}}
  (\href{http://dx.doi.org/10.1103/PhysRevA.79.053605}{\bibinfo{year}{2009}}).

\bibitem{campostrini_vicari_09}
\bibinfo{author}{M.~Campostrini} and \bibinfo{author}{E.~Vicari},
  \bibinfo{title}{{Critical Behavior and Scaling in Trapped Systems}},
  \bibinfo{journal}{\href{http://dx.doi.org/10.1103/PhysRevLett.102.240601}{Phys.
  Rev. Lett.}} \href{http://dx.doi.org/10.1103/PhysRevLett.102.240601}{{\bf
  \bibinfo{volume}{102}}, \bibinfo{pages}{240601}}
  (\href{http://dx.doi.org/10.1103/PhysRevLett.102.240601}{\bibinfo{year}{2009}}).

\bibitem{pollet_prokofev_10}
\bibinfo{author}{L.~Pollet}, \bibinfo{author}{N.~V. Prokof'ev}, and
  \bibinfo{author}{B.~V. Svistunov}, \bibinfo{title}{{Criticality in Trapped
  Atomic Systems}},
  \bibinfo{journal}{\href{http://dx.doi.org/10.1103/PhysRevLett.104.245705}{Phys.
  Rev. Lett.}} \href{http://dx.doi.org/10.1103/PhysRevLett.104.245705}{{\bf
  \bibinfo{volume}{104}}, \bibinfo{pages}{245705}}
  (\href{http://dx.doi.org/10.1103/PhysRevLett.104.245705}{\bibinfo{year}{2010}}).

\bibitem{Spielman_10}
\bibinfo{author}{K.~Jim\'enez-Garc\'{\i}a}, \bibinfo{author}{R.~L. Compton},
  \bibinfo{author}{Y.-J. Lin}, \bibinfo{author}{W.~D. Phillips},
  \bibinfo{author}{J.~V. Porto}, and \bibinfo{author}{I.~B. Spielman},
  \bibinfo{title}{{Phases of a Two-Dimensional {B}ose Gas in an Optical
  Lattice}},
  \bibinfo{journal}{\href{http://dx.doi.org/10.1103/PhysRevLett.105.110401}{Phys.
  Rev. Lett.}} \href{http://dx.doi.org/10.1103/PhysRevLett.105.110401}{{\bf
  \bibinfo{volume}{105}}, \bibinfo{pages}{110401}}
  (\href{http://dx.doi.org/10.1103/PhysRevLett.105.110401}{\bibinfo{year}{2010}}).

\bibitem{campostrini_vicari_10b}
\bibinfo{author}{M.~Campostrini} and \bibinfo{author}{E.~Vicari},
  \bibinfo{title}{Trap-size scaling in confined-particle systems at quantum
  transitions},
  \bibinfo{journal}{\href{http://dx.doi.org/10.1103/PhysRevA.81.023606}{Phys.
  Rev. A}} \href{http://dx.doi.org/10.1103/PhysRevA.81.023606}{{\bf
  \bibinfo{volume}{81}}, \bibinfo{pages}{023606}}
  (\href{http://dx.doi.org/10.1103/PhysRevA.81.023606}{\bibinfo{year}{2010}}).

\bibitem{campostrini_vicari_10c}
\bibinfo{author}{M.~Campostrini} and \bibinfo{author}{E.~Vicari},
  \bibinfo{title}{Quantum critical behavior and trap-size scaling of trapped
  bosons in a one-dimensional optical lattice},
  \bibinfo{journal}{\href{http://dx.doi.org/10.1103/PhysRevA.81.063614}{Phys.
  Rev. A}} \href{http://dx.doi.org/10.1103/PhysRevA.81.063614}{{\bf
  \bibinfo{volume}{81}}, \bibinfo{pages}{063614}}
  (\href{http://dx.doi.org/10.1103/PhysRevA.81.063614}{\bibinfo{year}{2010}}).

\bibitem{mahmud_duchon_11}
\bibinfo{author}{K.~W. Mahmud}, \bibinfo{author}{E.~N. Duchon},
  \bibinfo{author}{Y.~Kato}, \bibinfo{author}{N.~Kawashima},
  \bibinfo{author}{R.~T. Scalettar}, and \bibinfo{author}{N.~Trivedi},
  \bibinfo{title}{Finite-temperature study of bosons in a two-dimensional
  optical lattice},
  \bibinfo{journal}{\href{http://dx.doi.org/10.1103/PhysRevB.84.054302}{Phys.
  Rev. B}} \href{http://dx.doi.org/10.1103/PhysRevB.84.054302}{{\bf
  \bibinfo{volume}{84}}, \bibinfo{pages}{054302}}
  (\href{http://dx.doi.org/10.1103/PhysRevB.84.054302}{\bibinfo{year}{2011}}).

\bibitem{pollet_12}
\bibinfo{author}{L.~Pollet}, \bibinfo{title}{{Recent developments in quantum
  Monte Carlo simulations with applications for cold gases}},
  \bibinfo{journal}{\href{http://dx.doi.org/10.1088/0034-4885/75/9/094501}{Rep.
  Prog. Phys.}} \href{http://dx.doi.org/10.1088/0034-4885/75/9/094501}{{\bf
  \bibinfo{volume}{75}}, \bibinfo{pages}{094501}}
  (\href{http://dx.doi.org/10.1088/0034-4885/75/9/094501}{\bibinfo{year}{2012}}).

\bibitem{ceccarelli_torrero_13}
\bibinfo{author}{G.~Ceccarelli}, \bibinfo{author}{C.~Torrero}, and
  \bibinfo{author}{E.~Vicari}, \bibinfo{title}{Critical parameters from
  trap-size scaling in systems of trapped particles},
  \bibinfo{journal}{\href{http://dx.doi.org/10.1103/PhysRevB.87.024513}{Phys.
  Rev. B}} \href{http://dx.doi.org/10.1103/PhysRevB.87.024513}{{\bf
  \bibinfo{volume}{87}}, \bibinfo{pages}{024513}}
  (\href{http://dx.doi.org/10.1103/PhysRevB.87.024513}{\bibinfo{year}{2013}}).

\bibitem{angelone_campostrini_14}
\bibinfo{author}{A.~Angelone}, \bibinfo{author}{M.~Campostrini}, and
  \bibinfo{author}{E.~Vicari}, \bibinfo{title}{Universal quantum behavior of
  interacting fermions in one-dimensional traps: From few particles to the trap
  thermodynamic limit},
  \bibinfo{journal}{\href{http://dx.doi.org/10.1103/PhysRevA.89.023635}{Phys.
  Rev. A}} \href{http://dx.doi.org/10.1103/PhysRevA.89.023635}{{\bf
  \bibinfo{volume}{89}}, \bibinfo{pages}{023635}}
  (\href{http://dx.doi.org/10.1103/PhysRevA.89.023635}{\bibinfo{year}{2014}}).

\bibitem{zhang_18}
\bibinfo{author}{Y.~Zhang}, \bibinfo{author}{L.~Vidmar}, and
  \bibinfo{author}{M.~Rigol}, \bibinfo{title}{Information measures for a local
  quantum phase transition: Lattice fermions in a one-dimensional harmonic
  trap},
  \bibinfo{journal}{\href{http://dx.doi.org/10.1103/PhysRevA.97.023605}{Phys.
  Rev. A}} \href{http://dx.doi.org/10.1103/PhysRevA.97.023605}{{\bf
  \bibinfo{volume}{97}}, \bibinfo{pages}{023605}}
  (\href{http://dx.doi.org/10.1103/PhysRevA.97.023605}{\bibinfo{year}{2018}}).

\bibitem{jaksch_98}
\bibinfo{author}{D.~Jaksch}, \bibinfo{author}{C.~Bruder},
  \bibinfo{author}{J.~I. Cirac}, \bibinfo{author}{C.~W. Gardiner}, and
  \bibinfo{author}{P.~Zoller}, \bibinfo{title}{{Cold Bosonic Atoms in Optical
  Lattices}},
  \bibinfo{journal}{\href{http://dx.doi.org/10.1103/PhysRevLett.81.3108}{Phys.
  Rev. Lett.}} \href{http://dx.doi.org/10.1103/PhysRevLett.81.3108}{{\bf
  \bibinfo{volume}{81}}, \bibinfo{pages}{3108}}
  (\href{http://dx.doi.org/10.1103/PhysRevLett.81.3108}{\bibinfo{year}{1998}}).

\bibitem{greiner02}
\bibinfo{author}{M.~Greiner}, \bibinfo{author}{O.~Mandel},
  \bibinfo{author}{T.~Esslinger}, \bibinfo{author}{T.~H\"ansch}, and
  \bibinfo{author}{I.~Bloch}, \bibinfo{title}{{Quantum phase transition from a
  superfluid to a Mott insulator in a gas of ultracold atoms}},
  \bibinfo{journal}{\href{http://dx.doi.org/10.1038/415039a}{Nature}}
  \href{http://dx.doi.org/10.1038/415039a}{{\bf \bibinfo{volume}{415}},
  \bibinfo{pages}{39}}
  (\href{http://dx.doi.org/10.1038/415039a}{\bibinfo{year}{2002}}).

\bibitem{greiner_mandel_02b}
\bibinfo{author}{M.~Greiner}, \bibinfo{author}{O.~Mandel},
  \bibinfo{author}{T.~W. H\"ansch}, and \bibinfo{author}{I.~Bloch},
  \bibinfo{title}{Collapse and revival of the matter wave field of a
  $\mathrm{B}$ose-$\mathrm{E}$instein condensate},
  \bibinfo{journal}{\href{http://dx.doi.org/10.1038/nature00968}{Nature}}
  \href{http://dx.doi.org/10.1038/nature00968}{{\bf \bibinfo{volume}{419}},
  \bibinfo{pages}{51}}
  (\href{http://dx.doi.org/10.1038/nature00968}{\bibinfo{year}{2002}}).

\bibitem{bloch08}
\bibinfo{author}{I.~Bloch}, \bibinfo{author}{J.~Dalibard}, and
  \bibinfo{author}{W.~Zwerger}, \bibinfo{title}{Many-body physics with
  ultracold gases},
  \bibinfo{journal}{\href{http://dx.doi.org/10.1103/RevModPhys.80.885}{Rev.
  Mod. Phys.}} \href{http://dx.doi.org/10.1103/RevModPhys.80.885}{{\bf
  \bibinfo{volume}{80}}, \bibinfo{pages}{885}}
  (\href{http://dx.doi.org/10.1103/RevModPhys.80.885}{\bibinfo{year}{2008}}).

\bibitem{cazalilla_citro_review_11}
\bibinfo{author}{M.~A. Cazalilla}, \bibinfo{author}{R.~Citro},
  \bibinfo{author}{T.~Giamarchi}, \bibinfo{author}{E.~Orignac}, and
  \bibinfo{author}{M.~Rigol}, \bibinfo{title}{One dimensional bosons: From
  condensed matter systems to ultracold gases},
  \bibinfo{journal}{\href{http://dx.doi.org/10.1103/RevModPhys.83.1405}{Rev.
  Mod. Phys.}} \href{http://dx.doi.org/10.1103/RevModPhys.83.1405}{{\bf
  \bibinfo{volume}{83}}, \bibinfo{pages}{1405}}
  (\href{http://dx.doi.org/10.1103/RevModPhys.83.1405}{\bibinfo{year}{2011}}).

\bibitem{Fisher_1989}
\bibinfo{author}{M.~P.~A. Fisher}, \bibinfo{author}{P.~B. Weichman},
  \bibinfo{author}{G.~Grinstein}, and \bibinfo{author}{D.~S. Fisher},
  \bibinfo{title}{Boson localization and the superfluid-insulator transition},
  \bibinfo{journal}{\href{http://dx.doi.org/10.1103/PhysRevB.40.546}{Phys. Rev.
  B}} \href{http://dx.doi.org/10.1103/PhysRevB.40.546}{{\bf
  \bibinfo{volume}{40}}, \bibinfo{pages}{546}}
  (\href{http://dx.doi.org/10.1103/PhysRevB.40.546}{\bibinfo{year}{1989}}).

\bibitem{white_92}
\bibinfo{author}{S.~R. White}, \bibinfo{title}{Density matrix formulation for
  quantum renormalization groups},
  \bibinfo{journal}{\href{http://dx.doi.org/10.1103/PhysRevLett.69.2863}{Phys.
  Rev. Lett.}} \href{http://dx.doi.org/10.1103/PhysRevLett.69.2863}{{\bf
  \bibinfo{volume}{69}}, \bibinfo{pages}{2863}}
  (\href{http://dx.doi.org/10.1103/PhysRevLett.69.2863}{\bibinfo{year}{1992}}).

\bibitem{schollwoeck_05}
\bibinfo{author}{U.~Schollw\"ock}, \bibinfo{title}{The density-matrix
  renormalization group},
  \bibinfo{journal}{\href{http://dx.doi.org/10.1103/RevModPhys.77.259}{Rev.
  Mod. Phys.}} \href{http://dx.doi.org/10.1103/RevModPhys.77.259}{{\bf
  \bibinfo{volume}{77}}, \bibinfo{pages}{259}}
  (\href{http://dx.doi.org/10.1103/RevModPhys.77.259}{\bibinfo{year}{2005}}).

\bibitem{schollwoeck_11}
\bibinfo{author}{U.~Schollw\"ock}, \bibinfo{title}{The density-matrix
  renormalization group in the age of matrix product states},
  \bibinfo{journal}{\href{http://dx.doi.org/http://dx.doi.org/10.1016/j.aop.2010.09.012}{Annals
  of Physics}}
  \href{http://dx.doi.org/http://dx.doi.org/10.1016/j.aop.2010.09.012}{{\bf
  \bibinfo{volume}{326}}, \bibinfo{pages}{96 }}
  (\href{http://dx.doi.org/http://dx.doi.org/10.1016/j.aop.2010.09.012}{\bibinfo{year}{2011}}).

\bibitem{buonsante_07}
\bibinfo{author}{P.~Buonsante} and \bibinfo{author}{A.~Vezzani},
  \bibinfo{title}{{Ground-State Fidelity and Bipartite Entanglement in the
  Bose-Hubbard Model}},
  \bibinfo{journal}{\href{http://dx.doi.org/10.1103/PhysRevLett.98.110601}{Phys.
  Rev. Lett.}} \href{http://dx.doi.org/10.1103/PhysRevLett.98.110601}{{\bf
  \bibinfo{volume}{98}}, \bibinfo{pages}{110601}}
  (\href{http://dx.doi.org/10.1103/PhysRevLett.98.110601}{\bibinfo{year}{2007}}).

\bibitem{deng_11}
\bibinfo{author}{X.~Deng} and \bibinfo{author}{L.~Santos},
  \bibinfo{title}{{Entanglement spectrum of one-dimensional extended
  Bose-Hubbard models}},
  \bibinfo{journal}{\href{http://dx.doi.org/10.1103/PhysRevB.84.085138}{Phys.
  Rev. B}} \href{http://dx.doi.org/10.1103/PhysRevB.84.085138}{{\bf
  \bibinfo{volume}{84}}, \bibinfo{pages}{085138}}
  (\href{http://dx.doi.org/10.1103/PhysRevB.84.085138}{\bibinfo{year}{2011}}).

\bibitem{ejima_12}
\bibinfo{author}{S.~Ejima}, \bibinfo{author}{H.~Fehske},
  \bibinfo{author}{F.~Gebhard}, \bibinfo{author}{K.~zu~M\"unster},
  \bibinfo{author}{M.~Knap}, \bibinfo{author}{E.~Arrigoni}, and
  \bibinfo{author}{W.~von~der Linden}, \bibinfo{title}{{Characterization of
  Mott-insulating and superfluid phases in the one-dimensional Bose-Hubbard
  model}},
  \bibinfo{journal}{\href{http://dx.doi.org/10.1103/PhysRevA.85.053644}{Phys.
  Rev. A}} \href{http://dx.doi.org/10.1103/PhysRevA.85.053644}{{\bf
  \bibinfo{volume}{85}}, \bibinfo{pages}{053644}}
  (\href{http://dx.doi.org/10.1103/PhysRevA.85.053644}{\bibinfo{year}{2012}}).

\bibitem{pino_12}
\bibinfo{author}{M.~Pino}, \bibinfo{author}{J.~Prior}, \bibinfo{author}{A.~M.
  Somoza}, \bibinfo{author}{D.~Jaksch}, and \bibinfo{author}{S.~R. Clark},
  \bibinfo{title}{{Reentrance and entanglement in the one-dimensional
  Bose-Hubbard model}},
  \bibinfo{journal}{\href{http://dx.doi.org/10.1103/PhysRevA.86.023631}{Phys.
  Rev. A}} \href{http://dx.doi.org/10.1103/PhysRevA.86.023631}{{\bf
  \bibinfo{volume}{86}}, \bibinfo{pages}{023631}}
  (\href{http://dx.doi.org/10.1103/PhysRevA.86.023631}{\bibinfo{year}{2012}}).

\bibitem{alba_12}
\bibinfo{author}{V.~Alba}, \bibinfo{author}{M.~Haque}, and
  \bibinfo{author}{A.~M. L\"auchli}, \bibinfo{title}{{Boundary-Locality and
  Perturbative Structure of Entanglement Spectra in Gapped Systems}},
  \bibinfo{journal}{\href{http://dx.doi.org/10.1103/PhysRevLett.108.227201}{Phys.
  Rev. Lett.}} \href{http://dx.doi.org/10.1103/PhysRevLett.108.227201}{{\bf
  \bibinfo{volume}{108}}, \bibinfo{pages}{227201}}
  (\href{http://dx.doi.org/10.1103/PhysRevLett.108.227201}{\bibinfo{year}{2012}}).

\bibitem{alba_13}
\bibinfo{author}{V.~Alba}, \bibinfo{author}{M.~Haque}, and
  \bibinfo{author}{A.~M. L\"auchli}, \bibinfo{title}{{Entanglement Spectrum of
  the Two-Dimensional Bose-Hubbard Model}},
  \bibinfo{journal}{\href{http://dx.doi.org/10.1103/PhysRevLett.110.260403}{Phys.
  Rev. Lett.}} \href{http://dx.doi.org/10.1103/PhysRevLett.110.260403}{{\bf
  \bibinfo{volume}{110}}, \bibinfo{pages}{260403}}
  (\href{http://dx.doi.org/10.1103/PhysRevLett.110.260403}{\bibinfo{year}{2013}}).

\bibitem{frerot_16}
\bibinfo{author}{I.~Fr\'erot} and \bibinfo{author}{T.~Roscilde},
  \bibinfo{title}{{Entanglement Entropy across the Superfluid-Insulator
  Transition: A Signature of Bosonic Criticality}},
  \bibinfo{journal}{\href{http://dx.doi.org/10.1103/PhysRevLett.116.190401}{Phys.
  Rev. Lett.}} \href{http://dx.doi.org/10.1103/PhysRevLett.116.190401}{{\bf
  \bibinfo{volume}{116}}, \bibinfo{pages}{190401}}
  (\href{http://dx.doi.org/10.1103/PhysRevLett.116.190401}{\bibinfo{year}{2016}}).

\bibitem{deng_2013}
\bibinfo{author}{X.~Deng}, \bibinfo{author}{R.~Citro},
  \bibinfo{author}{E.~Orignac}, \bibinfo{author}{A.~Minguzzi}, and
  \bibinfo{author}{L.~Santos}, \bibinfo{title}{Bosonization and entanglement
  spectrum for one-dimensional polar bosons on disordered lattices},
  \bibinfo{journal}{\href{http://dx.doi.org/10.1088/1367-2630/15/4/045023}{New
  J. Phys.}} \href{http://dx.doi.org/10.1088/1367-2630/15/4/045023}{{\bf
  \bibinfo{volume}{15}}, \bibinfo{pages}{045023}}
  (\href{http://dx.doi.org/10.1088/1367-2630/15/4/045023}{\bibinfo{year}{2013}}).

\bibitem{Goldsborough_2015}
\bibinfo{author}{A.~M. Goldsborough} and \bibinfo{author}{R.~A. Römer},
  \bibinfo{title}{{Using entanglement to discern phases in the disordered
  one-dimensional Bose-Hubbard model}},
  \bibinfo{journal}{\href{http://dx.doi.org/10.1209/0295-5075/111/26004}{{EPL}
  (Europhysics Letters)}}
  \href{http://dx.doi.org/10.1209/0295-5075/111/26004}{{\bf
  \bibinfo{volume}{111}}, \bibinfo{pages}{26004}}
  (\href{http://dx.doi.org/10.1209/0295-5075/111/26004}{\bibinfo{year}{2015}}).

\bibitem{Campostrini_2010}
\bibinfo{author}{M.~Campostrini} and \bibinfo{author}{E.~Vicari},
  \bibinfo{title}{Scaling of bipartite entanglement in one-dimensional lattice
  systems with a trapping potential},
  \bibinfo{journal}{\href{http://dx.doi.org/10.1088/1742-5468/2010/08/p08020}{J.
  Stat. Mech.}} \href{http://dx.doi.org/10.1088/1742-5468/2010/08/p08020}{{\bf
  \bibinfo{volume}{{\rm (2010)}}}, \bibinfo{pages}{P08020}}.

\bibitem{Calabrese_2011}
\bibinfo{author}{P.~Calabrese}, \bibinfo{author}{M.~Mintchev}, and
  \bibinfo{author}{E.~Vicari}, \bibinfo{title}{Entanglement entropy of
  one-dimensional gases},
  \bibinfo{journal}{\href{http://dx.doi.org/10.1103/PhysRevLett.107.020601}{Phys.
  Rev. Lett.}} \href{http://dx.doi.org/10.1103/PhysRevLett.107.020601}{{\bf
  \bibinfo{volume}{107}}, \bibinfo{pages}{020601}}
  (\href{http://dx.doi.org/10.1103/PhysRevLett.107.020601}{\bibinfo{year}{2011}}).

\bibitem{vicari12}
\bibinfo{author}{E.~Vicari}, \bibinfo{title}{{Entanglement and particle
  correlations of Fermi gases in harmonic traps}},
  \bibinfo{journal}{\href{http://dx.doi.org/10.1103/PhysRevA.85.062104}{Phys.
  Rev. A}} \href{http://dx.doi.org/10.1103/PhysRevA.85.062104}{{\bf
  \bibinfo{volume}{85}}, \bibinfo{pages}{062104}}
  (\href{http://dx.doi.org/10.1103/PhysRevA.85.062104}{\bibinfo{year}{2012}}).

\bibitem{Calabrese_2015}
\bibinfo{author}{P.~Calabrese}, \bibinfo{author}{P.~Le~Doussal}, and
  \bibinfo{author}{S.~N. Majumdar}, \bibinfo{title}{Random matrices and
  entanglement entropy of trapped fermi gases},
  \bibinfo{journal}{\href{http://dx.doi.org/10.1103/PhysRevA.91.012303}{Phys.
  Rev. A}} \href{http://dx.doi.org/10.1103/PhysRevA.91.012303}{{\bf
  \bibinfo{volume}{91}}, \bibinfo{pages}{012303}}
  (\href{http://dx.doi.org/10.1103/PhysRevA.91.012303}{\bibinfo{year}{2015}}).

\bibitem{dubail_stephan_17}
\bibinfo{author}{J.~Dubail}, \bibinfo{author}{J.-M. Stéphan},
  \bibinfo{author}{J.~Viti}, and \bibinfo{author}{P.~Calabrese},
  \bibinfo{title}{{Conformal Field Theory for Inhomogeneous One-dimensional
  Quantum Systems: the Example of Non-Interacting Fermi Gases}},
  \bibinfo{journal}{\href{http://dx.doi.org/10.21468/SciPostPhys.2.1.002}{SciPost
  Phys.}} \href{http://dx.doi.org/10.21468/SciPostPhys.2.1.002}{{\bf
  \bibinfo{volume}{2}}, \bibinfo{pages}{002}}
  (\href{http://dx.doi.org/10.21468/SciPostPhys.2.1.002}{\bibinfo{year}{2017}}).

\bibitem{dubail_stephan_17b}
\bibinfo{author}{J.~Dubail}, \bibinfo{author}{J.-M. Stéphan}, and
  \bibinfo{author}{P.~Calabrese}, \bibinfo{title}{{Emergence of curved
  light-cones in a class of inhomogeneous Luttinger liquids}},
  \bibinfo{journal}{\href{http://dx.doi.org/10.21468/SciPostPhys.3.3.019}{SciPost
  Phys.}} \href{http://dx.doi.org/10.21468/SciPostPhys.3.3.019}{{\bf
  \bibinfo{volume}{3}}, \bibinfo{pages}{019}}
  (\href{http://dx.doi.org/10.21468/SciPostPhys.3.3.019}{\bibinfo{year}{2017}}).

\bibitem{eisler_bauernfeind_17}
\bibinfo{author}{V.~Eisler} and \bibinfo{author}{D.~Bauernfeind},
  \bibinfo{title}{{Front dynamics and entanglement in the XXZ chain with a
  gradient}},
  \bibinfo{journal}{\href{http://dx.doi.org/10.1103/PhysRevB.96.174301}{Phys.
  Rev. B}} \href{http://dx.doi.org/10.1103/PhysRevB.96.174301}{{\bf
  \bibinfo{volume}{96}}, \bibinfo{pages}{174301}}
  (\href{http://dx.doi.org/10.1103/PhysRevB.96.174301}{\bibinfo{year}{2017}}).

\bibitem{tonni_rodriguez_18}
\bibinfo{author}{E.~Tonni}, \bibinfo{author}{J.~Rodr{\'{\i}}guez-Laguna}, and
  \bibinfo{author}{G.~Sierra}, \bibinfo{title}{{Entanglement Hamiltonian and
  entanglement contour in inhomogeneous 1D critical systems}},
  \bibinfo{journal}{\href{http://dx.doi.org/10.1088/1742-5468/aab67d}{J. Stat.
  Mech.}} \href{http://dx.doi.org/10.1088/1742-5468/aab67d}{{\bf
  \bibinfo{volume}{{\rm (2018)}}}, \bibinfo{pages}{043105}}.

\bibitem{murciano_ruggiero_19}
\bibinfo{author}{S.~Murciano}, \bibinfo{author}{P.~Ruggiero}, and
  \bibinfo{author}{P.~Calabrese}, \bibinfo{title}{Entanglement and relative
  entropies for low-lying excited states in inhomogeneous one-dimensional
  quantum systems},
  \bibinfo{journal}{\href{http://dx.doi.org/10.1088/1742-5468/ab00ec}{J. Stat.
  Mech.}} \href{http://dx.doi.org/10.1088/1742-5468/ab00ec}{{\bf
  \bibinfo{volume}{{\rm (2019)}}}, \bibinfo{pages}{034001}}.

\bibitem{itensor}
{\em \bibinfo{title}{http://itensor.org}\/}.

\bibitem{carrasquilla_manmana_13}
\bibinfo{author}{J.~Carrasquilla}, \bibinfo{author}{S.~R. Manmana}, and
  \bibinfo{author}{M.~Rigol}, \bibinfo{title}{Scaling of the gap, fidelity
  susceptibility, and {B}loch oscillations across the
  superfluid-to-{M}ott-insulator transition in the one-dimensional
  {B}ose-{H}ubbard model},
  \bibinfo{journal}{\href{http://dx.doi.org/10.1103/PhysRevA.87.043606}{Phys.
  Rev. A}} \href{http://dx.doi.org/10.1103/PhysRevA.87.043606}{{\bf
  \bibinfo{volume}{87}}, \bibinfo{pages}{043606}}
  (\href{http://dx.doi.org/10.1103/PhysRevA.87.043606}{\bibinfo{year}{2013}}).

\bibitem{Ejima_2011}
\bibinfo{author}{S.~Ejima}, \bibinfo{author}{H.~Fehske}, and
  \bibinfo{author}{F.~Gebhard}, \bibinfo{title}{{Dynamic properties of the
  one-dimensional Bose-Hubbard model}},
  \bibinfo{journal}{\href{http://dx.doi.org/10.1209/0295-5075/93/30002}{{EPL}
  (Europhysics Letters)}}
  \href{http://dx.doi.org/10.1209/0295-5075/93/30002}{{\bf
  \bibinfo{volume}{93}}, \bibinfo{pages}{30002}}
  (\href{http://dx.doi.org/10.1209/0295-5075/93/30002}{\bibinfo{year}{2011}}).

\bibitem{sachdevbook}
\bibinfo{author}{S.~Sachdev}, {\em \bibinfo{title}{Quantum Phase
  Transitions}\/} (\bibinfo{publisher}{Cambridge University Press},
  \bibinfo{address}{New York}, \bibinfo{year}{2011}).

\bibitem{alba_comm}
{\em \bibinfo{title}{\rm V. Alba, private communication}\/}.

\end{thebibliography}

\end{document}